\documentclass[11pt]{article}
\usepackage[symbol]{footmisc}

\usepackage[left=2.5cm,top=2.5cm,right=2cm,bottom=2.5cm]{geometry}
\usepackage{amsmath, amssymb}

\usepackage{graphicx,subfigure,epsfig,epsf,color}
\usepackage{slashed}
\usepackage{feynmf}
\usepackage{graphicx}
\usepackage{graphics}
\usepackage{epstopdf}
\usepackage{subfigure}
\usepackage{amsfonts}
\usepackage{float}
\usepackage{sectsty}
\usepackage{hyperref}
\usepackage{cite}
\usepackage{pdflscape}
\usepackage{lipsum}
\begin{document}
	\date{}
	
	\title{Using entropy bounds to avoid the cosmological singularity and constrain cosmological particle production}
	\maketitle
	\begin{center}
		Hao Yu$~^{a,}$\footnote{yuhaocd@cqu.edu.cn},~Jin Li~$^{a,}$\footnote{cqujinli1983@cqu.edu.cn, corresponding author} \\
	\end{center}
	
	\begin{center}
		$^a$ Physics Department, Chongqing University, Chongqing 401331, China\\
	\end{center}
	\vspace*{0.25in}
	\begin{abstract}
		In this work, we study the applications of entropy bounds in two toy cosmological models with particle production (annihilation), i.e., a radiation-dominated universe and a dust-dominated universe. We consider the co-moving volume and the volume covered by the particle horizon of a given observer as the thermodynamic systems satisfying entropy bounds. For the Bekenstein bound and the spherical entropy bound, it is found that the cosmological singularity can be avoided and cosmological particle production needs to be truncated in some special cases. Our study can be extended to other cosmological models with particle production.
		
	\end{abstract}
	
	\section{Introduction}
Black hole thermodynamics~\cite{Bekenstein:1972tm,Bekenstein:1973ur,Hawking:1974rv,Hawking:1975vcx}, which leads to the formulation of the holographic principle~\cite{tHooft:1993dmi,Susskind:1994vu,Bousso:2002ju}, may be the key to the understanding of quantum gravity. In 1972, Bekenstein found that black holes could have entropy~\cite{Bekenstein:1972tm}, which triggered the research enthusiasm on black hole thermodynamics. In these studies, entropy is a crucial factor in the link between black holes and thermodynamics. However, for nearly a decade after Bekenstein found that black holes have entropy, few researchers considered the relationship between the entropy of black holes and the entropy of other gravitational systems. Until 1981, when Bekenstein studied the generalized second law (of thermodynamics) for a black hole~\cite{Bekenstein:1974ax}, he for the first time argued that the generalized second law implies the entropy of any weakly gravitating matter system in asymptotically flat space should satisfy a bound $S\leq  2\pi k E R/(\hbar c)$, where $E$ is the total mass-energy of the system and $R$ is the radius of the smallest sphere that fits around the system~\cite{Bekenstein:1980jp}. This entropy bound is also called the Bekenstein bound, which is independent of the gravitational theory. It is worth mentioning that Unruh and Wald did not agree with the original derivation of the Bekenstein bound~\cite{Unruh:1982ic,Unruh:1983ir}. They stated that the entropy bound of a black hole is not needed for the validity of the generalized second law if there exists buoyancy force of the thermal atmosphere near the black hole horizon. Subsequent studies show that, according to physical processes, gravitational theories, the background space-time, etc., entropy bounds may have different forms. In 1995, Susskind argued that applying the generalized second law to a transformation that a system is converted to a black hole, one can get the spherical entropy bound $S\leq k A/({4l_p^2})$, where $A$ is a properly defined area enclosing the system~\cite{Susskind:1994vu,Wald:1999vt}. Subsequently, inspired by the work of Fischler and Susskind~\cite{Fischler:1998st}, Bousso proposed a covariant entropy bound (which is also called Bousso bound) $S[L(B)]\leq A(B)/4$, where $A$ is the area of the boundary $B$~\cite{Bousso:1999xy}. The covariant entropy bound can be applied to any space-time including the strong gravitational system and satisfies general covariance, but it is only applicable to general relativity. In Ref.~\cite{Flanagan:1999jp}, the authors provided two ways to prove the covariant entropy bound and put forward a stronger entropy bound. In the year that the covariant entropy bound was proposed, Brustein and Veneziano proposed a causal entropy bound~\cite{Brustein:1999md}. The following year, Verlinde proposed the Bekenstein-Verlinde bound, the Bekenstein-Hawking bound and the Hubble bound~\cite{Verlinde:2000wg}. For more research on entropy bounds, one can refer to Refs.~\cite{Cai:2002bn, Brustein:2007hd,Ashtekar:2008gn,Casini:2008cr,Ali:2011ap,Dvali:2020wqi,Acquaviva:2020qbc,Buoninfante:2020guu}.
	
Although the concept of entropy bounds is the product of black hole research, it also has significant implications in cosmology. Some models of the Big Bang theory predict that there exists a cosmological singularity (initial singularity or Big Bang singularity) before the Big Bang, which contained all the energy and space-time of the universe. The Big Bang theory fits in well with cosmological observations and has been accepted by many physicists, but the initial singularity of the universe has been criticized. As a result, some cosmological models and theories have been proposed to explain or avoid the cosmological singularity, such as the cyclic model of the universe, multiverse, loop quantum gravity, etc. In 1989, Bekenstein found that the cosmological singularity is thermodynamically irrational~\cite{Bekenstein:1989wf} from the perspective of the entropy bound proposed by himself~\cite{Bekenstein:1980jp}. Recently, Powell et al. proposed a re-examination of Bekenstein's approach in a radiation-dominated universe and also verified that the Bekenstein bound can be a feasible alternative to avoiding the cosmological singularity~\cite{Powell:2020nzc}. In addition to the application of entropy bounds to the initial singularity of the universe, entropy bounds may also help us deduce the shape of the universe. For a closed universe, one can find that there exists a contradiction between the Fischler-Susskind bound and the positive curvature, which means that the Fischler-Susskind bound requires the shape of the universe to be non-closed~\cite{Fischler:1998st}. For more cosmological applications of various entropy bounds, one can refer to Refs.~\cite{Veneziano:1999ty,McInnes:2002gf,Ashtekar:2008gn,Brevik:2010jv,Lewandowski:2013aka,Bousso:2015eda,Banks:2018aed}.
	
It is an indisputable fact that the production and annihilation of particles occur continuously in our universe, so the cosmological model with particle production (annihilation) is more consistent with the real universe. The mechanisms and applications of particle production in cosmology have long attracted the attention of physicists (see a recent review~\cite{Ford:2021syk} and references therein). In the 1960s, Parker established the first micromechanism of particle production in the context of cosmology with the quantum field theory in curved space-time~\cite{Parker:1968mv,Parker:1969au,Parker:1972kp,Ford:1978ip}. Then, particle production in different background space-time has been investigated successively~\cite{Zeldovich:1971mw,Mottola:1984ar,Kuzmin:1998kk,Dunne:2005sx}. A few years ago, Harko found that, from a thermodynamic point of view, gravitational induced particle production can happen in non-minimal coupling theories~\cite{Harko:2014pqa,Harko:2015pma}, which provides a new mechanism for cosmological particle production. At present, most of the studies suggest that the production and annihilation of particles in an expanding universe have a profound theoretical foundation and they can not be ignored in the universe, especially in the early period. As for the applications of particle production in cosmology, they include multiple aspects, such as avoiding the cosmological singularity~\cite{Prigogine:1986zz,Prigogine:1988zz,Prigogine:1989zz}, explaining entropy production of the universe~\cite{Prigogine:1986zz,Prigogine:1988zz,Prigogine:1989zz,Kolb1990,SolaPeracaula:2019kfm}, accelerating the expansion of the universe~\cite{Traschen:1990sw,Prigogine:1989zz,Zimdahl:1999tn,Nunes:2015rea,Nunes:2016aup,Pan:2016bug}, and triggering inflation~\cite{Ford:1986sy,Abramo:1996ip}.

In this work, we study entropy bounds in the cosmological model with particle production (annihilation). We use entropy bounds to judge whether the cosmological singularity is thermodynamically rational in a radiation-dominated universe and a dust-dominated universe. Moreover, entropy bounds may be able to constrain cosmological particle production, which is a new application of entropy bounds in cosmology. The constraint on cosmological particle production is of significance for cosmology because it could provide us with truncations for some interactions in the context of cosmology. In the past, we usually used the (generalized) second law of thermodynamics to constrain cosmological particle production and the results were not unsatisfactory because it can only constrain the sign of the particle production rate~\cite{Yu:2018qzl,SolaPeracaula:2019kfm}.
	
The paper is organized as follows. Sec.~\ref{sec2} is a brief review on cosmological particle production and the corresponding entropy. In Sec.~\ref{sec3}, we discuss the Bekenstein bound and the spherical entropy bound in a radiation-dominated universe with particle production. We focus on the cosmological singularity and the constraint on cosmological particle production inside the co-moving volume and the volume covered by the particle horizon of a given observer. Then, in Sec.~\ref{sec4}, we study similar content in a dust-dominated universe with particle production. The last part, Sec.~\ref{sec5}, is conclusions and discussions.
	
	\section{Cosmological particle production and entropy}
	\label{sec2}
	In this section, we discuss cosmological particle production and entropy in the context of a homogeneous and isotropic universe, which can be described by the Friedmann-Lemat${\hat\i}$re-Robertson-Walker metric:
	\begin{eqnarray}
		\text{d}s^2=-\text{d}t^2+a^2(t)\left(\frac{\text{d}r^2}{1-\tilde k\,r^2}+r^2\text{d}\theta^2+r^2\text{sin}^2\theta \,\text{d}\phi^2\right).
	\end{eqnarray}
	For simplicity, we consider that the universe is spatially flat, i.e., $\tilde k=0$. Here, $a(t)$ is the scale factor. In this work, all the components of the universe are regarded as ideal fluids, so the energy-momentum tensor of all the ideal fluids is given as
	\begin{eqnarray}
		T_{\mu\nu}=(p+\rho)u_{\mu}u_{\nu}+p\,g_{\mu\nu},
	\end{eqnarray}
	where  $\rho$ and $p$ represent the total energy density and pressure of the ideal fluids, respectively. The four-velocity of the ideal fluids satisfies $u_{\nu}u^{\nu}=-1$. For general relativity, the Friedmann equations are given by
	\begin{eqnarray}
		H^2\!\!&=&\!\!\frac{8\pi G}{3}\rho,\\
		\dot H+H^2\!\!&=&\!\!-\frac{4\pi G}{3}\left(\rho+3{p}\right).
	\end{eqnarray}
Now, we focus on a spherical system with radius $R$ in the universe. According to the purpose of research, the physical meaning of $R$ can be multiple, such as the scale factor, the particle horizon, the apparent horizon, the radius of the visible universe, etc. Assuming that there are $N$ particles in the system, one can define the particle production rate as
	\begin{eqnarray}\label{5}
		\Gamma=\frac{\text d N}{\text d t}\frac{1}{N}.
	\end{eqnarray}
Usually, if $R$ is the scale factor and there is no interaction between these particles and other matter (or the background space-time), we have $\Gamma=0$. For other cases of $R$, even if there is no interaction, $\Gamma$ is generally nonzero due to the evolution of $R$. If the particle number in the system is non-conserved, the entropy of the system will be affected by the production (increase) or annihilation (decrease) of particles. We label the current particle number as $N_0$ and set the current radius of the system to $R_0$. If the entropy of these particles is extensive\footnote{In this work, we do not consider the non-extensive statistical entropy, such as Tsallis entropy~\cite{Tsallis:1987eu}. Some studies indicate that, compared to classical statistics, non-extensive statistics may be more applicable to a gravitational system~\cite{Tsallis:1987eu,Wilk:1999dr,Tsallis:2012js}.} in a homogeneous and isotropic gravitational system (such as the cosmological model we study), then the entropy of the system at time $t$ can be written as
	\begin{eqnarray}
		S_t=s(n_t)\,R_t^3,
	\end{eqnarray}
	where $n_t={N_t}/{V_t}={N_t}/{R_t^3}$ is the particle number density and $s(n_t)$ is the entropy per volume at time $t$. The radius $R_t$ of the system at time $t$ is dependent on the way the universe is expanding. The particle number $N_t$ at time $t$ is given by
	\begin{eqnarray}
		N_t=N_0\,\text{exp}\left[\int_{t_0}^{t}\Gamma\,\text d t \right].
	\end{eqnarray}
	
If the system is isolated, one can constrain the evolution of the system by the second law of thermodynamics:
	\begin{eqnarray}\label{8}
		\frac{\text d S_t}{\text d t}=\frac{\text d s(n_t)}{\text d n}\left(\Gamma\,n_t-3\frac{n_t}{R_t}\frac{\text d R_t}{\text d t}\right)+3s(n_t)\,R_t^2\frac{\text d R_t}{\text d t}>0.
	\end{eqnarray}
	On the other hand, if the system possesses area entropy (such as the system covered by the apparent horizon~\cite{Bousso:2002ju,BakRey2000,Bousso1999,Bousso2005,CaiKim2005,AkbarCai2006,CaiCao2007,Faraoni2011,
		FaraoniBook2015,Melia2018}), then the system can be constrained by the generalized second law of thermodynamics~\cite{Bekenstein:1972tm,Bekenstein:1973ur,Bekenstein:1974ax}:
	\begin{eqnarray}\label{9}
		\frac{\text d S_t}{\text d t}+	\frac{\text d S_A}{\text d t}>0.
	\end{eqnarray}
The area entropy $S_A$ of the system is usually proportional to the area of the system, but the coefficient is related to the gravitational theory. Owing to inequality~(\ref{8}) [or inequality~(\ref{9})], the particle production rate $\Gamma$ in the universe can not be arbitrary. Therefore, the (generalized) second law of thermodynamics is a means of constraining cosmological particle production (or the interaction between different substances in the context of cosmology).
	
However, for a general radius $R$, the system is not isolated, so it may be not appropriate to use the (generalized) second law of thermodynamics to constrain cosmological particle production. In this work, we omit the area entropy of the system and try to constrain cosmological particle production with entropy bounds. Since entropy bounds do not require the corresponding system to be closed, we can study cosmological particle production in any system. On the other hand, we will also examine the effect of cosmological particle production on the cosmological singularity in the view of entropy bounds~\footnote{In this work, we are not concerned about the mechanism by which matter was produced (annihilated) at the beginning of the universe.}. The entropy bounds employed in this work are the Bekenstein bound~\cite{Bekenstein:1980jp} and the spherical entropy bound~\cite{Susskind:1994vu,Wald:1999vt,Bousso:2002ju}. In order to obtain analytical solutions of the Friedmann equations, each of the toy cosmological models we consider only contains a species of matter.

\section{Entropy bounds and particle production in a radiation-dominated universe}
	\label{sec3}
In 1989, Bekenstein studied the particle horizon of a given observer in a radiation-dominated universe and he found the cosmological singularity is thermodynamically impossible by considering the Bekenstein bound~\cite{Bekenstein:1989wf}. In Bekenstein's Friedmann model, there is no cosmological particle production inside the co-moving volume, which may be inappropriate for the early universe because non-relativistic particles were not decoupled with other matter in the early universe.  In this section, we extend Bekenstein's Friedmann model by introducing particle production into the radiation-dominated universe and discuss the relationship among the cosmological singularity, particle production and entropy bounds.
	
As a toy model, we simplify the radiation-dominated universe as a universe containing only photons. The production of photons can be ascribed to the coupling between photons and the background space-time~\cite{Harko:2014pqa,Harko:2015pma} (or the running vacuum~\cite{Sola:2013gha,Sola:2014tta,SolaPeracaula:2019kfm}). Since we do not know how to define the entropy of the background space-time, the total entropy of the universe is considered to be only dependent on photons. We assume that the number of photons per unit volume is
	\begin{eqnarray}
		n=\frac{2k^3\zeta(3)}{\pi^2c^3\hbar^3}T^3,
	\end{eqnarray}
where $c$ is the light speed, $k$ is the Boltzmann constant, $\zeta(\mathfrak{n})$ is the Riemann zeta function. If there is no coupling between photons and the background space-time (i.e., there is no particle production in the  co-moving volume), the temperature $T$ of photons should be proportional to $a^{-1}$. For a spherical system with radius $R$ in the universe, with Eq.~(\ref{5}) and $N=n\,R^3$, one can obtain the production rate of photons in the system is given as
	\begin{eqnarray}\label{11}
		\Gamma=\frac{1}{n}\frac{\text d n}{\text d t}+\frac{1}{R^3}\frac{\text d R^3}{\text d t}=3\left(\frac{\dot T}{T}+\frac{\dot R}{R}\right).
	\end{eqnarray}
	Since the entropy of photons per unit volume is given by
	\begin{eqnarray}
		s=\frac{4\pi^2k^4}{45c^3\hbar^3}T^3,
	\end{eqnarray}
	the entropy of photons inside the spherical system with radius $R$ is
	\begin{eqnarray}\label{13}
		S=s\,R^3=\frac{4\pi^2k^4}{45c^3\hbar^3}T^3 R^3=\frac{2\pi^4k}{45\zeta(3)}N=\frac{2\pi^4k}{45\zeta(3)}N_0\,\text{exp}\left[\int_{t_0}^{t}\Gamma\,\text d t \right],
	\end{eqnarray}
	where $N_0$ is the current number of photons in the system. When $R$ and $\Gamma$ are given, we can calculate the entropy of photons in the system at any time. Next, we consider that $R$ is the scale factor (i.e., the system is the co-moving volume) or the particle horizon of a given observer (i.e., the system is the volume covered by the particle horizon). As $t$ approaches to zero, the production (annihilation) of photons may influence the cosmological singularity from the perspective of entropy bounds. Moreover, as $t$ increases, entropy bounds may require that there exists a truncation for the production of photons.

	\subsection{Co-moving volume}
	\label{sec31}
	If $R$ is the scale factor, according to Eqs.~(\ref{11}) and~(\ref{13}), the entropy of photons inside the co-moving volume is given as
	\begin{eqnarray}\label{14}
		S=\frac{2\pi^4k}{45\zeta(3)}N_0\,\text{exp}\left[3\int_{t_0}^{t}\left(\frac{\dot T}{T}+\frac{\dot a}{a}\right)\,\text d t \right].
	\end{eqnarray}
	If $\Gamma=0$, then $S$ is a constant.
	
	We first consider the Bekenstein bound, which requires the entropy of any weakly gravitating matter system to satisfy~\cite{Bekenstein:1980jp}
	\begin{eqnarray}\label{15}
		S\leq \frac{2\pi k}{\hbar c} E R,
	\end{eqnarray}
where $E$ is the total mass-energy including any rest mass and $R$ is the radius of a sphere that can enclose the given system. For the photons inside the co-moving volume, we have $E=\frac{\pi^2k^4}{15c^3\hbar^3}a^3T^4$, so the Bekenstein bound can be expressed as
	\begin{eqnarray}\label{16}
		\frac{2\pi^4k}{45\zeta(3)}N_0\,\text{exp}\left[3\int_{t_0}^{t}\left(\frac{\dot T}{T}+\frac{\dot a}{a}\right)\,\text d t \right]\leq \frac{2\pi k}{\hbar c} \frac{\pi^2k^4}{15c^3\hbar^3}a^4T^4.
	\end{eqnarray}

When $\Gamma=0$, the energy density of photons inside the co-moving volume is given by $\rho=\rho_0\,a^4$ (the Friedmann equations) or $\rho=\frac{\pi^2k^4}{15c^3\hbar^3}T^4$ (thermodynamic state functions for black-body photons), so $aT=a_0 T_0$ is a constant. Taking $N_0=\frac{2k^3\zeta(3)}{\pi^2c^3\hbar^3}a_0^3T_0^3$ into inequality~(\ref{16}), then the Bekenstein bound can be simplified as
	\begin{eqnarray}\label{17}
		\frac 23\leq \frac{\pi k}{\hbar c} a_0T_0,
	\end{eqnarray}
where $k\sim1.38*10^{-23}$ J/K and $\hbar c\sim3.16*10^{-26}$ J$\cdot$m. If we set $a_0=1$ (m) and $T_0\sim2.7$ K, then this inequality is tenable. Therefore, for a radiation-dominated universe without particle production, if its current temperature is the same as our real universe, applying the Bekenstein bound to the co-moving volume can not avoid the cosmological singularity. Moreover, based on inequality~(\ref{17}), one can find that if $T_0<2*10^{-4}$ K, the entropy of photons inside the co-moving volume will conflict with the Bekenstein bound.
	
	When $\Gamma\neq0$, $a\,T$ will evolve over time. We can rewrite Eq.~(\ref{11}) as
	\begin{eqnarray}
		\Gamma=3\frac{\text d (aT)}{\text d t}\frac{1}{aT}.
	\end{eqnarray}
	Then, one can obtain
	\begin{eqnarray}\label{19}
		\text{exp}\left[\int_{t_0}^{t}\Gamma\,\text d t \right]=\left(\frac{aT}{a_0T_0}\right)^3.
	\end{eqnarray}
	Substituting it into inequality (\ref{16}) to cancel $aT$ and $N_0$ yields
	\begin{eqnarray}\label{20a}
		\frac23\leq \frac{\pi k}{\hbar c} a_0T_0\,\text{exp}\left[\frac13\int_{t_0}^{t}\Gamma\,\text d t \right],	
	\end{eqnarray}
where we can set $a_0=1$ (m) and $T_0\sim2.7$ K.
	
	For $0\leq t<t_0$ and $\Gamma>0$, there must be a critical time $t_c$ corresponding to the equality sign of inequality~(\ref{20a}). When $t<t_c$, inequality~(\ref{20a}) will be violated, which means that the Bekenstein bound requires the initial time of the universe to be nonzero. However, due to the existence of $\Gamma$, the scale factor does not satisfy $a(t=0)=0$, so the nonzero initial time is not equivalent to the nonzero initial volume. In the later discussion, we will see that, for $\Gamma>0$, the universe naturally has no initial singularity according to the solution the scale factor. Therefore, in this case we do not need the Bekenstein bound to avoid the cosmological singularity. We will focus on the effect of the Bekenstein bound on the initial volume of the universe. Regarding this, we can see detailed examples later.
	
	For $0\leq t<t_0$ and $\Gamma\leq0$, since  the right-hand side of inequality~(\ref{20a}) increases with the decrease of $t$, the Bekenstein bound does not help to avoid the cosmological singularity.
	
	For $t_0\leq t$ and $\Gamma\geq0$, the right-hand side of inequality~(\ref{20a}) increases with $t$. Based on the previous analysis, the Bekenstein bound is always true. Therefore, the photons inside the co-moving volume can be continuously produced and so the particle production rate $\Gamma$ can not be limited by the Bekenstein bound.
	
	For $t_0\leq t$ and $\Gamma<0$, as $t$ increases, the entropy of photons inside the co-moving volume will violate the Bekenstein bound. Therefore, photons in the universe can not keep annihilating and the interaction between photons and the background space-time must be truncated at some point. It is worth noting that the result seems to defy our physical intuition. With the increase of the scale factor and the annihilation of photons, the entropy and energy of photons inside the co-moving volume will be reduced synchronously. However, the reduction of the energy is faster than that of the entropy, so the critical condition of the Bekenstein bound (i.e., $S=\frac{2\pi k}{\hbar c} E R$) will appear and then be broken as $t$ increases. It is different from the critical condition of the Bekenstein bound for compact objects, which only occurs in the case of the black hole and will not be violated as the mass of the black hole changes. The difference stems from the selection of the volume of the system. The former is the co-moving volume of the universe, which is affected by the background space-time. But, the latter is the volume of a sphere that can enclose the given system, which is related to the total mass-energy of the system. With regard to the issue, it is not the topic of this work, so we do not discuss it any further here.
	
	Next, we consider the spherical entropy bound, which is given as
	\begin{eqnarray}\label{21}
		S\leq \frac{k A}{4l_p^2}.
	\end{eqnarray}
	Here, $A$ is the area of the system and $l_p$ is the Planck length. For the photons inside the co-moving volume ($A=a^2$), the spherical entropy bound can be expressed as
	\begin{eqnarray}\label{22}
		\frac{2\pi^4k}{45\zeta(3)}N_0\,\text{exp}\left[\int_{t_0}^{t}\Gamma\,\text d t \right]\leq \frac{k \,a^2}{4l_p^2}.
	\end{eqnarray}
	
	When $\Gamma=0$, since the scale factor is monotonically increasing, once $a^2\geq \frac{8\pi^4l_p^2}{45\zeta(3)}N_0$, the spherical entropy bound will not be violated. Therefore,  $a=\left(\frac{8\pi^4l_p^2}{45\zeta(3)}N_0\right)^{1/2}$ is the lower bound of the scale factor, which means that the spherical entropy bound could avoid the cosmological singularity in a radiation-dominated universe without particle production.
	
	When $\Gamma\neq0$, using Eq.~(\ref{19}) to eliminate $\Gamma$, we can obtain
	\begin{eqnarray}\label{23}
		\frac{2\pi^4k}{45\zeta(3)}N_0\left(\frac{aT}{a_0T_0}\right)^3\leq \frac{k \,a^2}{4l_p^2}.
	\end{eqnarray}
	In order to judge whether this inequality is true, we need to figure out the relationship between $a$ and $T$ by solving Eq.~(\ref{19}). Here, we can discuss some issues qualitatively in the absence of the solutions of Eq.~(\ref{19}). Reviewing inequality~(\ref{22}), one can find that if $\Gamma=\frac2a\frac{\text d a}{\text d t}>0$, we have $\text{exp}\left[\int_{t_0}^{t}\Gamma\,\text d t \right]=\left(\frac{aT}{a_0T_0}\right)^3=\frac{a^2}{a_0^2}$. Taking $\left(\frac{aT}{a_0T_0}\right)^3=\frac{a^2}{a_0^2}$ and $N_0=\frac{2k^3\zeta(3)}{\pi^2c^3\hbar^3}a_0^3T_0^3$ into inequality (\ref{23}) yields
	\begin{eqnarray}\label{25}
		\frac{4k^3\pi^2}{45c^3\hbar^3}{a_0^3T_0^3}\leq \frac{a_0^2}{4l_p^2}.
	\end{eqnarray}
	If $a_0=1$ (m) and $T_0\sim2.7$ K, this inequality is true and independent of the evolution of the scale factor. We can set $\Gamma_c=\frac2a\frac{\text d a}{\text d t}$ as a critical particle production rate. Note that $\Gamma_c$ is a function of time not a point.
	
	When $\Gamma>\Gamma_c$, we have $\left(\frac{aT}{a_0T_0}\right)^3>\frac{a^2}{a_0^2}$ for $t>t_0$. As long as $t$ is large enough, inequality (\ref{22}) will be violated. Therefore, to meet the spherical entropy bound there should be a truncation for the production of photons at some point. Moreover, for $\Gamma>\Gamma_c>0$, as we mentioned earlier, there is naturally no cosmological singularity due to the solution of the scale factor, but the spherical entropy bound might modify the initial volume of the universe.
	
	When $\Gamma<\Gamma_c$ and $t>t_0$, we have $\left(\frac{aT}{a_0T_0}\right)^3<\frac{a^2}{a_0^2}$ and then inequality (\ref{22}) must be true. Therefore, the spherical entropy bound can not constrain the production (annihilation) rate of photons. But, if $\Gamma<\Gamma_c$ and $t<t_0$, we have $\left(\frac{aT}{a_0T_0}\right)^3>\frac{a^2}{a_0^2}$ and then there could exist a lower bound for the scale factor, which can avoid the cosmological singularity.
	
	In this section, we have not provided the solutions of Eq.~(\ref{19}), so there is only qualitative analysis. Next, we will discuss some similar issues in detail in the presence of analytical solutions.

	\subsection{Particle horizon}
	\label{sec32}
	
	We will now take the volume covered by the particle horizon of a given observer as the thermodynamic system satisfying entropy bounds. For convenience, the production rate of photons inside the particle horizon will be characterized by the production rate of photons inside the co-moving volume. In an identical universe, the difference between the two particle production rates depends on the evolution of the universe. Therefore, the subsequent calculations are also characterized by the production rate of photons inside the co-moving volume.
	
	We start from the solution of the energy density of photons in the presence of particle production. According to the laws of thermodynamics for open systems, the energy density of photons satisfies\cite{Prigogine:1986zz,Prigogine:1988zz,Prigogine:1989zz}
	\begin{eqnarray}\label{26}
		\text d (\rho a^3)+p \,\text d a^3=\frac{\rho+p}{n}\text d (na^3).
	\end{eqnarray}
	With $\Gamma=\frac{\text d N}{\text d t}\frac{1}{N}$, $N=n a^3$, and $p=\frac13\rho$, we have
	\begin{eqnarray}\label{27}
		\frac{\text d \rho}{\text d a}+\frac{4}{a}\rho-\frac{4}{3}\rho\,\Gamma \frac{\text d t}{\text d a}=0.
	\end{eqnarray}
	Therefore, the solution of the energy density of photons can be expressed as
	\begin{eqnarray}\label{28}
		\rho=\frac{\rho_0\,a_0^4}{a^4}\,\text{exp}\left[\frac43\int_{t_0}^{t}\Gamma\,\text d t \right].
	\end{eqnarray}
	For $\Gamma=0$ and $a_0=1$, it degenerates into the standard solution $\rho=\rho_0\, a^{-4}$.
	
\begin{figure}[h]
		\centering
		\renewcommand{\figurename}{Fig.}
		\includegraphics[width=12cm,height=8cm]{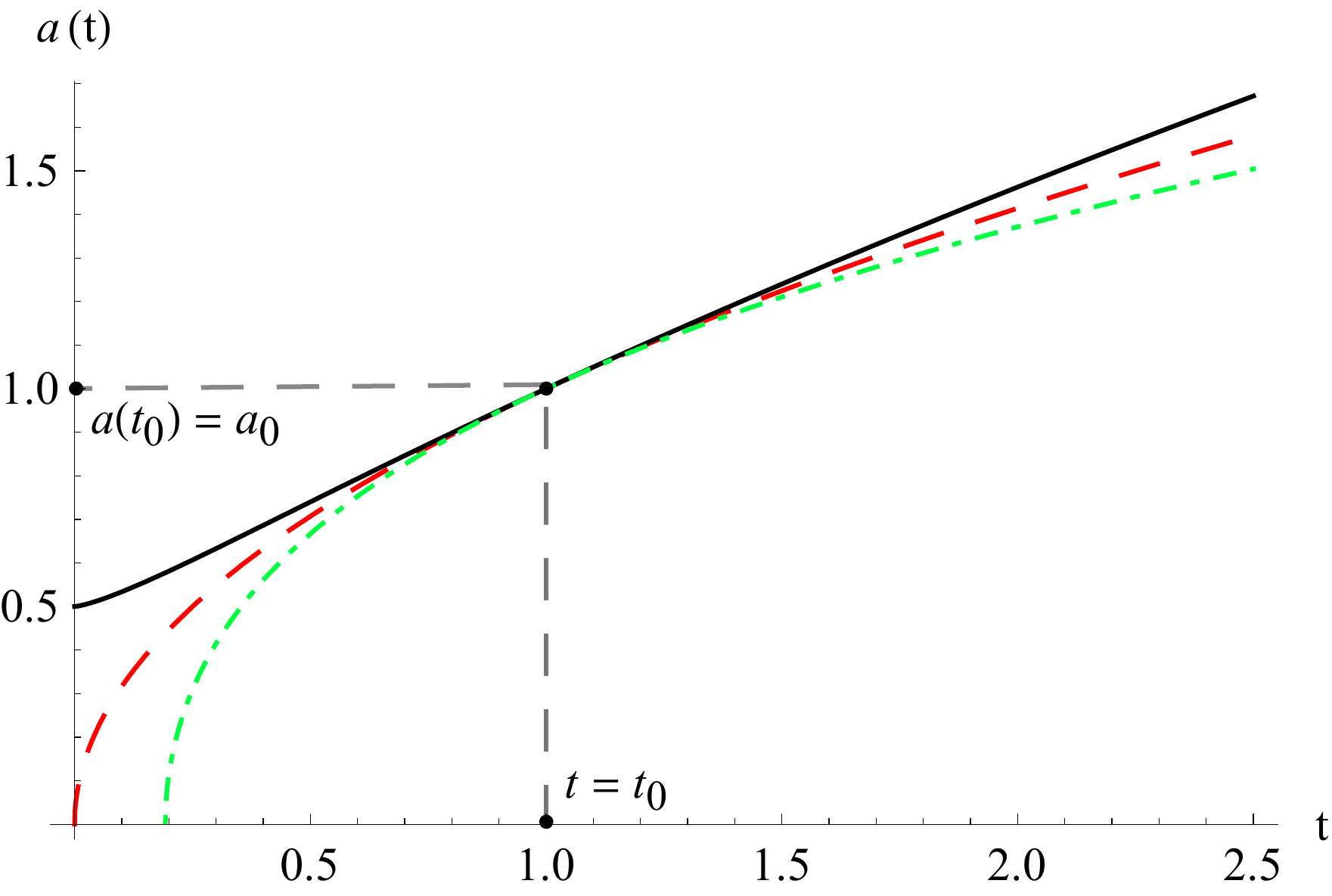}
		\caption{Plot of the scale factor (\ref{32}). In the schematic diagram, we have set $a(t_0)=a_0=1$, $t_0=1$, and $b_1=1$ with $g=0$. We study three sets of $g$: $g=\frac{1}2$ ($\Gamma>0$, $b_1=\frac34$ and $d_1=\frac14$) marked with the black solid line, $g=0$ ($\Gamma=0$, $b_1=1$ and $d_1=0$) marked with the red dashed line, and $g=-\frac{1}2$ ($\Gamma<0$, $b_1=\frac32$ and $d_1=-\frac12$) marked with the green dash-dotted line, respectively. These three lines intersect at the point $(t_0,a_0)$. When $\Gamma>0$ and $t=0$, the scale factor is larger than zero. When $\Gamma<0$, the minimum value of time is larger than zero.} 	\label{F1}		
	\end{figure}

	Since the universe is dominated by photons, we can ignore other matter when we solve the Friedmann equations\footnote{Note that the reason why the number of photons inside the co-moving volume is not conserved is the interaction between photons and other matter (or the background space-time). Therefore, ignoring the matter (or the background space-time) interacting with photons is a rough approximation. However, as a toy model, we actually only need an analytical solution of the scale factor increasing monotonically over time. Incorporating the matter (or the background space-time) interacting with photons into the Friedmann equations will only make it more difficult to solve the equations analytically and is not helpful for the following research.}. Thus, taking Eq.~(\ref{28}) into the Friedmann equations, we finally get
	\begin{eqnarray}\label{29}
		a^2-a_0^2=2(G_0\rho_0)^{\frac12}a_0^2\int_{t_0}^{t}\text{exp}\left[\frac23\int_{t_0}^{t'}\Gamma\,\text d t \right]\text d t',
	\end{eqnarray}
	where $G_0=\frac{8}{3}\pi G$ with $G$ the Newtonian constant. When $\Gamma=0$, we have $a\sim t^{1/2}$. As for $\Gamma\neq0$, to get the analytical solution of the scale factor, we have to presuppose a specific form of $\Gamma$. We can analyze qualitatively the properties of the solution of the scale factor for different $\Gamma$'s. If $\Gamma=0$ results in $a=b\, t^{1/2}$ (where $a(t_0)=a_0$ and $a(0)=0$), it can be expected that $\Gamma>0$ corresponds to $a>b\, t^{1/2}$ (where $a(t_0)=a_0$ and $a(0)>0$) and $\Gamma<0$ corresponds to $a<b\, t^{1/2}$ (where $a(t_0)=a_0$ and $a(0)<0$). Note that $a(0)>0$ (i.e., $\Gamma>0$) means that the beginning ($t=0$) of the universe is not a singularity~\cite{Prigogine:1986zz,Prigogine:1988zz,Prigogine:1989zz}. Moreover, $a(0)<0$ (i.e., $\Gamma<0$) indicates that the singularity (i.e., $a(t)=0$) of the universe appears at the non-beginning ($t\neq0$) of the universe (see Fig.~\ref{F1}). In this section, we must be cautious about these two strange situations because the particle horizon may be unnormal, which should be avoided. Next, we discuss these issues in detail.
	
	The particle horizon of a given observer in the radiation-dominated universe is given as
	\begin{eqnarray}\label{30}
		R_H=\int_{t_s}^t\frac{c}{a(t')}\text d t',
	\end{eqnarray}
	where $t_s\geq0$ is the time at which the observer starts observing at $r=0$. Since $a(t)>0$, $R_H$ is a monotonically increasing function of $t$. For $a=b\, t^{1/2}$, we have $R_H=2b^{-1}c\,\left(t^{1/2}-t_s^{1/2}\right)$. In Ref.~\cite{Bekenstein:1989wf}, based on the particle horizon of a given observer in a radiation-dominated universe and the Bekenstein bound, Bekenstein pointed out that the cosmological singularity is not thermodynamically possible (see Refs.~\cite{Bekenstein:1989wf,Powell:2020nzc} for more details).
	
	Now, let us analyze how $\Gamma$ affects the cosmological singularity. For a system covered by the particle horizon of a given observer, the Bekenstein bound can be expressed as
	\begin{eqnarray}\label{31}
		\frac{4\pi^2k^4}{45c^3\hbar^3}T^3 R_H^3\leq \frac{2\pi k}{\hbar c} \frac{\pi^2k^4}{15c^3\hbar^3}T^4 R_H^4.
	\end{eqnarray}	
	When $\Gamma=0$, there is a lower bound for the scale of the universe determined by the Bekenstein bound~\cite{Bekenstein:1989wf,Powell:2020nzc}. Since $\Gamma$ influences the solution of the scale factor (which determines the particle horizon and the temperature of photons), it may modify the lower bound of the scale of the universe. From the previous analysis, a positive-definite $\Gamma$ will lead to a bigger scale factor (the scale factor given by Eq.~(\ref{29}) should satisfy $a>b\,t^{1/2}$) and a smaller particle horizon $R_H<2b^{-1}c\,\left(t^{1/2}-t_s^{1/2}\right)$. However, a positive-definite $\Gamma$ will also cause the temperature of photons to be higher. Therefore, we can not easily estimate the effect of $\Gamma$ on the lower bound of the scale of the universe on the basis of inequality~(\ref{31}). With a specific $\Gamma$, we can analyze the issues further.
	
	We set $\Gamma=\frac g t$, where $g>0$ ($g<0$) represents particle production (annihilation). Then, taking it into Eq.~(\ref{29}), the solution of the scale factor can be expressed as
	\begin{eqnarray}\label{32}
		a^2=b_1t^{\frac23g+1}+d_1,
	\end{eqnarray}
	where $b_1=2(G_0\rho_0)^{\frac12}\frac{3a_0^2}{3+2g}t_0^{-\frac{2g}{3}}>0$ and $d_1=-2(G_0\rho_0)^{\frac12}\frac{3a_0^2}{3+2g}t_0+a_0^2$. We consider that $g>0$, $g=0$, and $g<0$ correspond to $d_1>0$, $d_1=0$, and $d_1<0$, respectively. From Fig.~\ref{F1}, one can find that when $\Gamma< 0$, the value of the scale factor at $t=0$ will be negative by extending the green dash-dotted line, which may lead to confusion when we calculate the particle horizon. In order to avoid such a nuisance, the lower bound of the integral in Eq.~(\ref{30}) should not be smaller than the minimum value of time (which ensures that the scale factor is non-negative).
	
	Then, the particle horizon can be expressed as
	\begin{eqnarray}\label{33}
		R_H\!\!\!&=&\!\!\!\frac{c}{d_1} t \sqrt{b_1 t^{\frac{2 g}{3}+1}+d_1}\, _2F_1\left(1,\frac{2 g+9}{4 g+6};\frac{2 g+6}{2 g+3};-\frac{b_1 }{d_1}t^{\frac{2 g}{3}+1}\right)\nonumber\\
		\!\!\!&-&\!\!\!\frac{c}{d_1} t_s \sqrt{b_1 t_s^{\frac{2 g}{3}+1}+d_1}\, _2F_1\left(1,\frac{2 g+9}{4 g+6};\frac{2 g+6}{2 g+3};-\frac{b_1 }{d_1}t_s^{\frac{2 g}{3}+1}\right),
	\end{eqnarray}
	where $_2F_1(a,b,c,z)$ is a hypergeometric function. For convenience, we can assume that, for any $\Gamma$, we always have  $F(t)=\int\frac{c}{a(t)}\text d t$ and $F(t_s)=0$, which will not alter our subsequent analysis and conclusion. Then, the second line in the above equation is vanishing.

	With $\Gamma=\frac g t$, $\rho=\frac{\pi^2k^4}{15c^3\hbar^3}T^4$, Eqs.~(\ref{28}) and (\ref{31}), the temperature of photons is given as
	\begin{eqnarray}\label{34}
		T^4=\frac{15c^3\hbar^3 \rho_0 a_0^4}{ \pi^2k^4 \left(b_1t^{\frac23g+1}+d_1\right)^2 }\left(\frac{t}{t_0}\right)^{4g/3}.
	\end{eqnarray}
	Reviewing inequality~(\ref{31}), to study the influence of $\Gamma$ on the cosmological singularity, we only need to figure out the impact of the parameter $g$ on $TR_H$. Labelling all the positive coefficients in $TR_H$ as a positive-definite parameter $M_0$, $TR_H$ can be written as
	\begin{eqnarray}\label{35}
		TR_H=M_0\frac{t^{\frac{g}{3}+1}}{d_1\,t_0^{\frac g3}} \, _2F_1\left(1,\frac{2 g+9}{4 g+6};\frac{2 g+6}{2 g+3};-\frac{b_1 }{d_1}t^{\frac{2 g}{3}+1}\right),
	\end{eqnarray}
	where $M_0=c\left(\frac{15c^3\hbar^3\rho_0 a_0^4}{\pi^2k^4}\right)^{1/4}$. For a given $g$, $TR_H$ generally can be simplified. However, since the hypergeometric function is complicated, for a given $g$, it is more reasonable to calculate $TR_H$ starting from Eq.~(\ref{32}). Next, we consider the three cases of $g$ adopted in Fig.~\ref{F1}.
	
\begin{figure}[h]
		\centering
		\renewcommand{\figurename}{Fig.}
		\includegraphics[width=8cm]{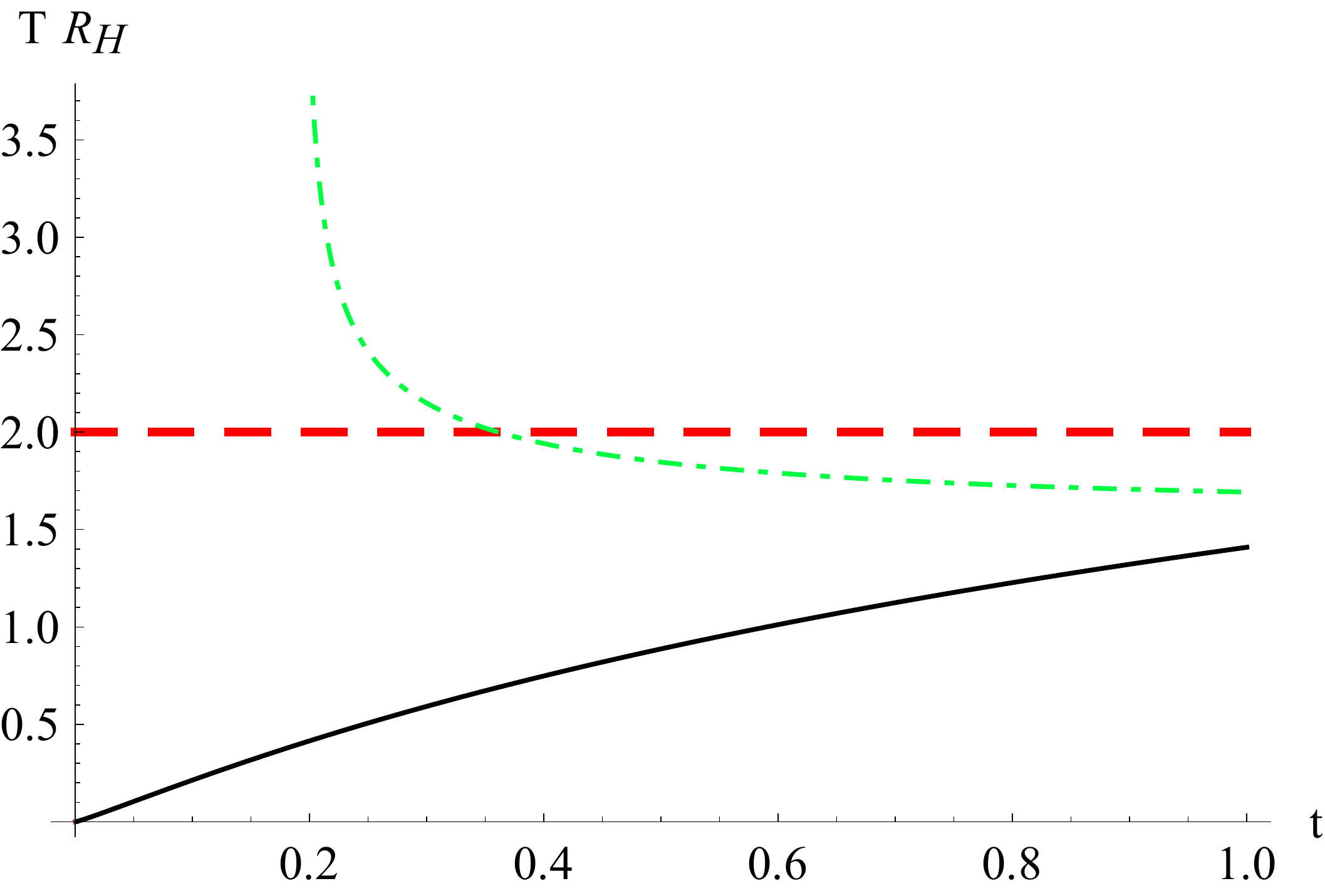}
		\includegraphics[width=8cm]{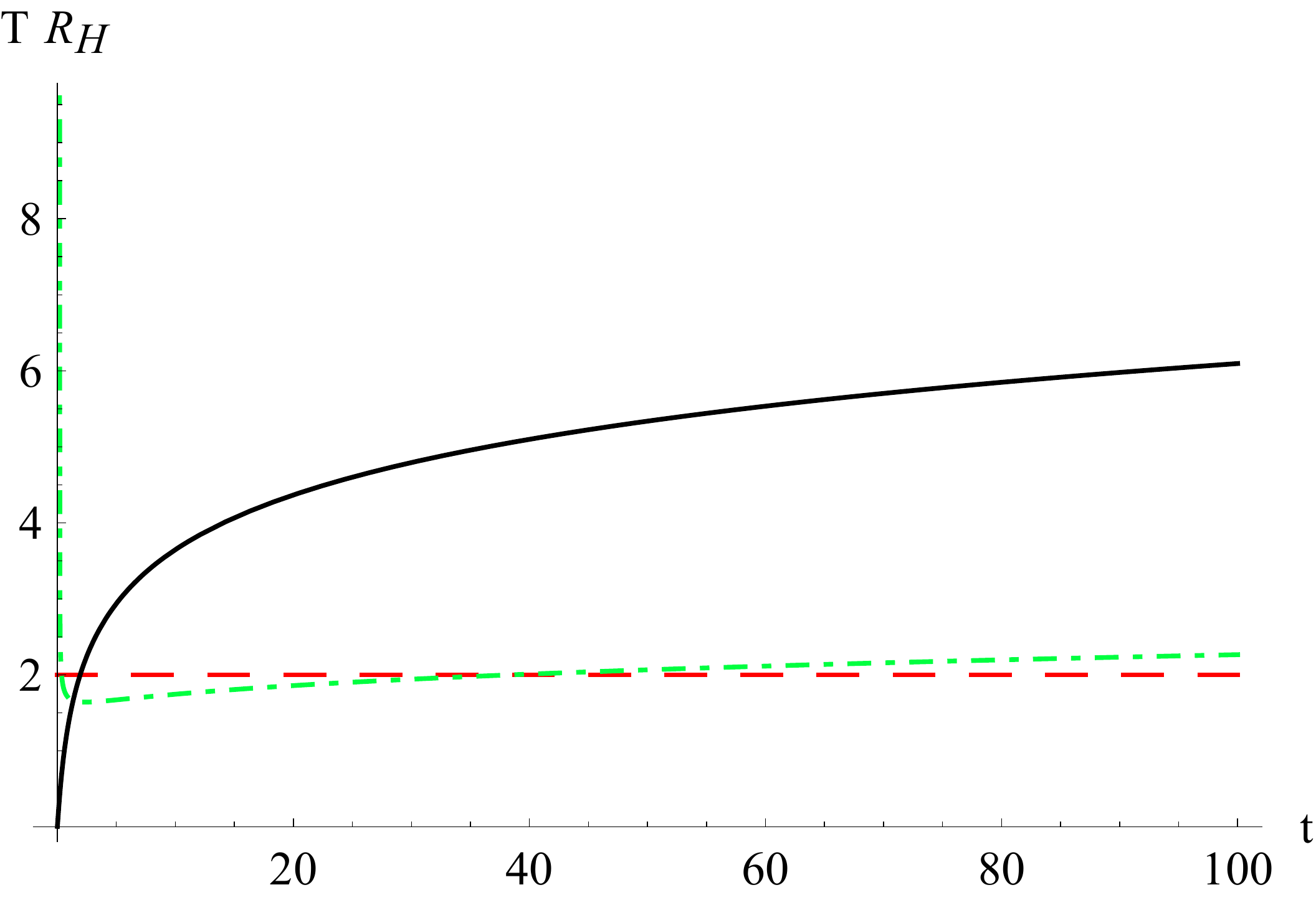}
		\caption{Plot of $T R_H$ (\ref{35}). In the schematic diagram, we have set $t_0=1$, $M_0=1$ and $b_1=1$ with $g=0$. We study three sets of $g$: $g=\frac{1}2$ ($\Gamma>0$, $b_1=\frac34$ and $d_1=\frac14$) marked with the black solid line, $g=0$ ($\Gamma=0$, $b_1=1$ and $d_1=0$) marked with the red dashed line, and $g=-\frac{1}2$ ($\Gamma<0$, $b_1=\frac32$ and $d_1=-\frac12$) marked with the green dash-dotted line, respectively.} \label{F2}		
	\end{figure}

	When $\Gamma=0$ and $t_s=0$, $TR_H=2 M_0 b_1^{-\frac32}$ is a constant (see the red dashed line in Fig.~\ref{F2}). In this case, one can prove that inequality~(\ref{31}) is tenable if $a_0=1$ (m) and $T_0\sim2.7$ K, so the cosmological singularity can not be avoided. Note that $t_s=0$ just corresponds to a special observer. If there are other observers who deny the rationality of the cosmological singularity, the universe should not have the initial singularity. When $\Gamma=0$ and $t_s>0$, one can find that $TR_H=2 M_0 b_1^{-\frac32}\left(1-t_s^{\frac12}\cdot t^{-\frac12}\right)$ increases with $t$ and has a maximum value. When $t\rightarrow t_s$, we have $TR_H\rightarrow0$, which inevitably causes inequality~(\ref{31}) to fail. Therefore, for a given $t_s$, $t$ must satisfy
	\begin{eqnarray}\label{355}
	t\geq\left(1-\frac{\hbar c}{3\pi k\,M_0} b_1^{\frac32}\right)^{-2} t_s>0.
    \end{eqnarray}
	We set $t_c$ as the critical time corresponding to the equality sign of the above inequality. Then, $a(t_c)$ characterizes the critical scale (i.e., lower bound of the scale) of the universe determined by the Bekenstein bound. If $t_s$ can approach to zero, the critical value (i.e., lower bound) of the scale factor can also approach to zero. However, if there is a minimum scale for time (such as the Planck-Wheeler time), then $a(t_c)$ can not approach to zero~\cite{Bekenstein:1989wf,Powell:2020nzc}. We will not dig and delve what value $t_s$ should take here. We just show that the Bekenstein bound could require a (nonzero) critical value for the scale factor in the radiation-dominated universe without particle production ($\Gamma=0$), so the cosmological singularity can be avoided.
	
	When $\Gamma\neq0$, in order to study the effect of $\Gamma$ on the critical value of the scale factor, we first analyze the critical time for $\Gamma\neq0$. For convenience, we set $t_s$ as the minimum value of time (see Fig.~\ref{F1}). Note that here the minimum value of time depends on the solution of the scale factor and has no connection with the Bekenstein bound. The ``minimum value of time'' determined by the Bekenstein bound is called critical time. We plot the schematic diagram of $T R_H$ in Fig.~\ref{F2}. From the left panel in Fig.~\ref{F2}, we can find that when $t\ll t_0$, if there exists production of photons, $TR_H$ (see the black solid line) will be smaller than the one in the case of $\Gamma=0$ (see the red dashed line). According to inequality~(\ref{31}), if the critical time for $\Gamma=0$ is $t_c$, then the critical time for $\Gamma>0$ would be $t_{c1}>t_c$. Moreover, from Fig.~\ref{F1}, the scale factor for $\Gamma>0$ is larger than that for $\Gamma=0$, so $t_{c1}>t_c$ means that the critical value of the scale factor for $\Gamma>0$ is larger than that for $\Gamma=0$. In other words, the production of photons causes the critical scale of the universe to be larger. As we mentioned earlier, for $\Gamma>0$, the solution of the scale factor naturally removes the cosmological singularity (i.e., $a(t=0)>0$, see Fig.~\ref{F1}). Here, the lower bound of the scale factor determined by the Bekenstein bound is unquestionably larger than $a(t=0)$. As for $\Gamma<0$, it is difficult to tell what the critical time is, because it could be $t_{c2}<t_c$ or may not exist. The former means that the critical value of the scale factor for $\Gamma<0$ is smaller than that for $\Gamma=0$, and the latter means the cosmological singularity can appear.
	
	Next, we analyze the constraint of the Bekenstein bound on the production of photons. One can find, from the right panel in Fig.~\ref{F2}, that with the increase of $t$, whatever the value of $\Gamma$ is, $TR_H$ is always increasing. Only when $\Gamma=0$ and $t_s=0$, $TR_H$ is a constant, for which the Bekenstein bound is tenable. Recalling inequality~(\ref{31}), the larger $TR_H$, the easier it is satisfied. Since $TR_H$ increases with $t$, the Bekenstein bound usually will not be violated in the late universe. Therefore, when we consider the entropy of photons in a system covered by the particle horizon in the late radiation-dominated universe, the Bekenstein bound can not provide effective constraint on the production of photons (i.e., the Bekenstein bound is easily satisfied). The interaction between photons and other matter (or the background space-time) does not need to be truncated.
	
	Now, we consider the spherical entropy bound for a system covered by the particle horizon of a given observer. We still assume that $a_0=1$ (m), $T_0\sim2.7$ K, and $t_s$ is the minimum value of time. With Eq.~(\ref{28}) and $\rho=\frac{\pi^2k^4}{15c^3\hbar^3}T^4$, the spherical entropy bound can be written as
	\begin{eqnarray}\label{36}
		S=\frac{4\pi^2k^4}{45c^3\hbar^3}T^3 R_H^3=\frac{M_0^3}{c^3}\text{exp}\left[\int_{t_0}^{t}\Gamma\,\text d t \right]\frac{ R_H^3}{a^3}	
		\leq \frac{k R_H^2}{4l_p^2},
	\end{eqnarray}
	where $M_0=c\left(\frac{15c^3\hbar^3\rho_0 a_0^4}{\pi^2k^4}\right)^{1/4}$.
	
\begin{figure}[h]
		\centering
		\renewcommand{\figurename}{Fig.}
		\includegraphics[width=12cm,height=8cm]{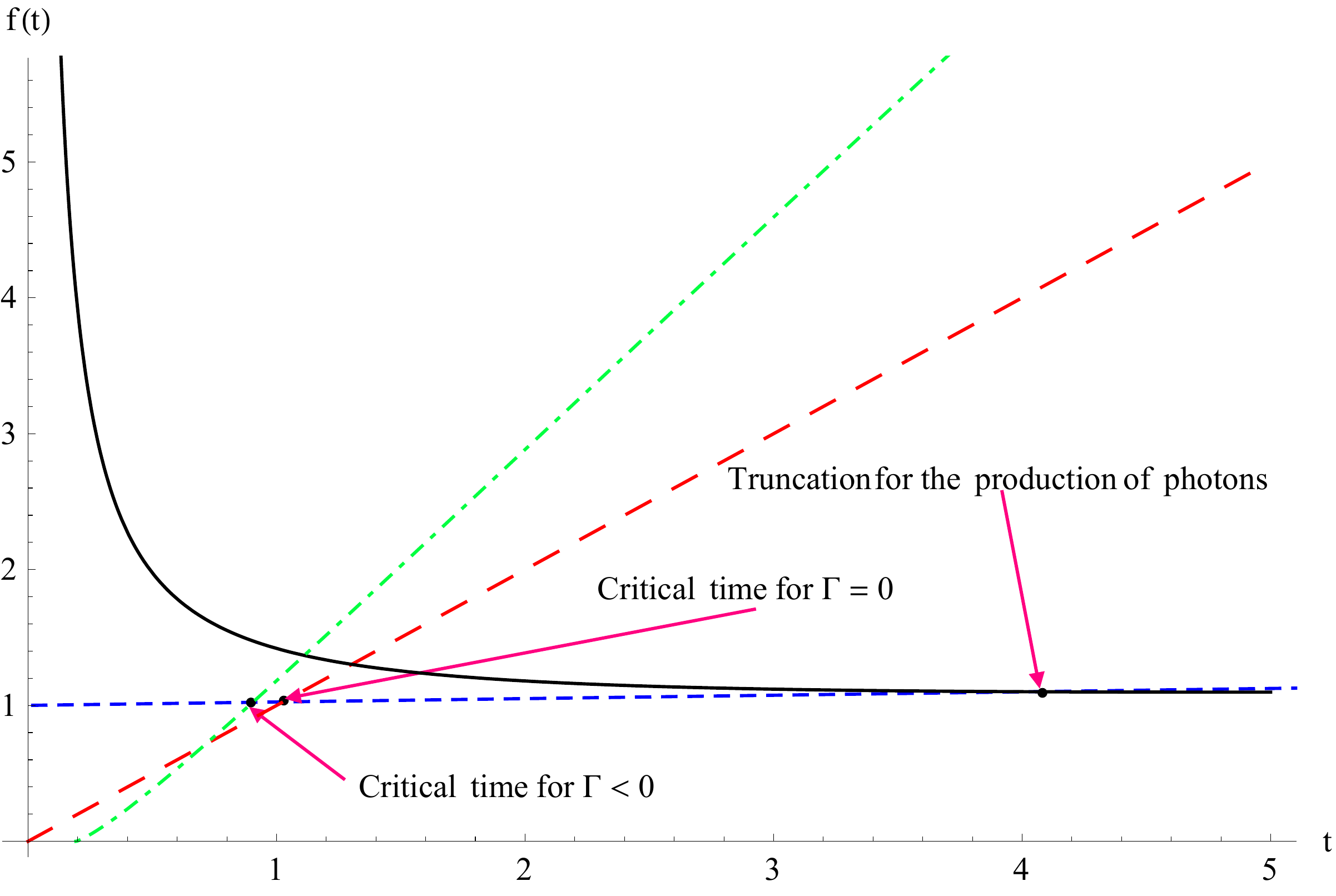}
		\caption{Plot of $f(t)$. There are three sets of $f(t)$: $f(t)=\left(\frac{t}{t_0}\right)^{-1/2}
			\frac{ R_H^{-1}}{\left(\frac34t^{4/3}+\frac14\right)^{-3/2}}$, $f(t)=t$, and  $f(t)=\left(\frac{t}{t_0}\right)^{1/2}
			\frac{ R_H^{-1}}{\left(\frac32t^{2/3}-\frac12\right)^{-3/2}}$. They correspond to	
			$g=\frac{1}2$ ($\Gamma>0$, $b_1=\frac34$ and $d_1=\frac14$) marked with the black solid line, $g=0$ ($\Gamma=0$, $b_1=1$ and $d_1=0$) marked with the red dashed line, and $g=-\frac{1}2$ ($\Gamma<0$, $b_1=\frac32$ and $d_1=-\frac12$) marked with the green dash-dotted line, respectively.} \label{F4}		
	\end{figure}

	When $\Gamma=0$, considering $a=t^{1/2}$ and $R_H=2c\,\left(t^{1/2}-t_s^{1/2}\right)$, inequality~(\ref{36}) can be simplified as
	\begin{eqnarray}\label{37}
		\frac{c^3}{M_0^3}\frac{k}{8c\,l_p^2} \left(1-\frac{t_s^{\frac12}}{t^{\frac12}}\right)^{-1} t
		\geq 1.
	\end{eqnarray}
	Since $t> t_s\geq 0$, for a given nonzero $t_s$, when $t\rightarrow t_s$, the left-hand side of the above inequality will approach to infinity. As $t$ increases, it reaches the minimum at $t=\frac94 t_s$ and then monotonically increases to infinity as $t\rightarrow \infty$. In this case, it is not easy to determine the lower bound of time (critical time) according to inequality~(\ref{37}). However, when $t_s=0$, we can find that inequality~(\ref{37}) requires a nonzero lower bound for time (see the red dashed line in Fig.~\ref{F4}). As long as there exists an observer requiring a nonzero lower bound for time, the lower bound of time must be nonzero. Therefore, considering the spherical entropy bound in a system covered by the particle horizon in the radiation-dominated universe, the cosmological singularity can be avoided.
	
	When $\Gamma\neq0$, we can still set $\Gamma=\frac{g}t$. Based on the previous calculations, if $g=\frac12$, inequality (\ref{36}) can be reduced as
	\begin{eqnarray}\label{38}
		\frac{c^3}{M_0^3} \frac{k}{4l_p^2}
		\left(\frac{t}{t_0}\right)^{-1/2}
		\frac{ R_H^{-1}}{\left(\frac34t^{4/3}+\frac14\right)^{-3/2}}	
		\geq 1,
	\end{eqnarray}
	where $R_H$ is given by Eq.~(\ref{33}) with $t_s=0$ (the minimum value of time, see Fig.~\ref{F1}). If $g=-\frac12$, inequality (\ref{36}) is reduced as
	\begin{eqnarray}\label{39}
		\frac{c^3}{M_0^3} \frac{k}{4l_p^2}
		\left(\frac{t}{t_0}\right)^{1/2}
		\frac{ R_H^{-1}}{\left(\frac32t^{2/3}-\frac12\right)^{-3/2}}	
		\geq 1,
	\end{eqnarray}
	where $R_H$ is given by Eq.~(\ref{33}) with $t_s=3^{-3/2}$ (the minimum value of time, see Fig.~\ref{F1}).
	
	In order to compare the critical times (the lower bounds of time) for the three cases, we can set $\frac{c^3}{M_0^3}\frac{k}{8c\,l_p^2}=1$ and $t_0=1$. Then, we label uniformly the left-hand sides of these three inequalities as a function $f(t)$. The evolutions of $f(t)$ over time for the three cases are plotted in Fig.~\ref{F4}. From Fig.~\ref{F4}, one can find that when there exists production of photons ($\Gamma>0$, see the black solid line), $f(t)$ decreases monotonically with time, so the corresponding critical time does not exist. In other words, even if $t\rightarrow0$, inequality (\ref{38}) still holds. Therefore, in this case the spherical entropy bound can not modify the critical scale of the universe (which depends on the solution of the scale factor). On the other hand, since $f(t)$ decreases monotonically with time, there should be a truncation for the production of photons (see the black solid line). When there exists annihilation of photons ($\Gamma<0$, see the green dash-dotted line), $f(t)$ increases monotonically with time. We find that the critical time for such case is smaller than the one obtained by $\Gamma=0$. According to Fig.~\ref{F1}, the scale factor for $\Gamma<0$ is smaller than that for $\Gamma=0$, so the critical scale of the universe for $\Gamma<0$ must be smaller than that for $\Gamma=0$. Moreover, since $f(t)$ increases monotonically with time, the spherical entropy bound can not constrain the annihilation of photons in the radiation-dominated universe. It is worth mentioning that these results are based on $t_s$ being the minimum value of time. As for other $t_s$'s, there may be different results, which we will not continue to discuss in this work.

	\section{Entropy bounds and particle production in a dust-dominated universe}
	\label{sec4}
	In the late 1980s, Prigogine proposed that cosmological particle production can avoid the initial singularity and solve the entropy problem~\cite{Prigogine:1986zz,Prigogine:1988zz,Prigogine:1989zz}. In his cosmological model, the entropy of the universe is expressed as the product of the particle number and the specific entropy (the average entropy of a single particle). If there exists particle production in the universe, the entropy inside the co-moving volume will keep increasing and so the entropy problem can be solved. In this section, we consider a similar toy cosmological model, i.e., a dust-dominated universe. The entropy of dust in any system can be expressed as
	\begin{eqnarray}\label{339}
		S=\sigma(t) N=\sigma(t)N_0\,\text{exp}\left[\int_{t_0}^{t}\Gamma\,\text d t \right],
	\end{eqnarray}
	where $\sigma(t)$ is the specific entropy of dust and $\Gamma$ is the production rate of dust in the system. Next, we study the effect of the production of dust on the cosmological singularity in light of entropy bounds and the constraint of entropy bounds on the production of dust. We still focus on the co-moving volume and the volume covered by the particle horizon of a given observer.

	\subsection{Co-moving volume}
	\label{sec41}
	For the co-moving volume, we have $N_0=n_0 a_0^3$ in Eq.~(\ref{339}) and $\Gamma$ is the production rate of dust inside the co-moving volume. Comparing Eq.~(\ref{339}) with Eq.~(\ref{14}), one can find that if $\sigma(t)$ is a constant, the entropy evolution of the dust-dominated universe is similar to that of the radiation-dominated universe. Therefore, an evolving $\sigma(t)$ has more general properties in characterizing the entropy evolution of the universe. Moreover, the solutions of the Friedmann equations for the dust-dominated universe are different from the ones for the radiation-dominated universe, which could also lead to different conclusions for similar situations.
	
	We first consider the Bekenstein bound, which, for the dust inside the co-moving volume, is given as
	\begin{eqnarray}\label{340}
		S=\sigma(t)N_0\,\text{exp}\left[\int_{t_0}^{t}\Gamma\,\text d t \right]\leq \frac{2\pi k}{\hbar c} E\, a.
	\end{eqnarray}
	For convenience, the potential energy of dust can be integrated into the rest mass of a single particle. In this case, the energy of dust inside the co-moving volume can be written as
	\begin{eqnarray}
	E=N\cdot m c^2=N_0\,\text{exp}\left[\int_{t_0}^{t}\Gamma\,\text d t \right]\cdot m c^2,
		\end{eqnarray}
	where $m$ is the equivalent mass of a single particle. The lower bound of $m$ corresponds to the rest mass of a single particle. Since the temperature of the universe is changing continuously, $m$ also varies with time. Theoretically, as $a\rightarrow0$, $m$ could be infinite. So the Bekenstein bound is reduced as
	\begin{eqnarray}\label{341}
		\sigma(t)\leq \frac{2\pi k c}{\hbar } m\, a.
	\end{eqnarray}
	
	If $\sigma(t)$ is a constant, the Bekenstein bound must be robust in the late dust-dominated universe due to the nonzero lower bound of $m$. Therefore, the production (annihilation) of dust in the late universe will not be limited by the Bekenstein bound. In addition, whether there is a lower bound for the scale factor relies on the evolutions of $m$, $\sigma(t)$ and the scale factor itself. If inequality~(\ref{341}) does not hold when $a$ is less than a certain threshold, then the threshold is the lower bound of the scale factor determined by the Bekenstein bound, which avoids the cosmological singularity. But, if inequality~(\ref{341}) holds for any $a$, the cosmological singularity can not be avoided.
	
	For a general $\sigma(t)$, since it can be independent of the scale factor, it is convenient to absorb $m$ into $\sigma(t)$ as a new parameter. Next, we set $\sigma_m(t)=\sigma(t)/m$ signifying the entropy of dust per unit mass, so inequality~(\ref{341}) can be re-expressed as
	\begin{eqnarray}\label{342}
		\sigma_m(t)\leq \frac{2\pi k c}{\hbar } \, a.
	\end{eqnarray}
	Note that although there is no apparent $\Gamma$ in the inequality, the solution of the scale factor depends on $\Gamma$. Therefore, $\Gamma$ and $\sigma_m(t)$ jointly determine the cosmological singularity, and $\Gamma$ can be constrained by the Bekenstein bound for a given $\sigma_m(t)$. With an analytical solution of the scale factor, we can discuss these issues in detail.
	
	Since we have absorbed the potential energy of dust into its rest mass, the (equivalent) pressure of dust can be approximately equal to zero and the (equivalent) energy density of dust satisfies
	\begin{eqnarray}\label{343}
		\frac{\text d \rho}{\text d a}+\frac{3}{a}\rho-\rho\,\Gamma \frac{\text d t}{\text d a}=0.
	\end{eqnarray}
	The solution of the above equation can be expressed as
	\begin{eqnarray}\label{344}
		\rho=\frac{\rho_0a_0^3}{a^3}\,\text{exp}\left[\int_{t_0}^{t}\Gamma\,\text d t \right].
	\end{eqnarray}
	Taking it into the Friedmann equations, one can get the solution of the scale factor:
	\begin{eqnarray}\label{345}
		a^{\frac32}-a_0^{\frac32}=\frac32(G_0\rho_0)^{\frac12}a_0^{\frac32}\int_{t_0}^{t}\text{exp}\left[\frac12\int_{t_0}^{t'}\Gamma\,\text d t \right]\text d t'.
	\end{eqnarray}
	
	Reviewing inequality~(\ref{342}), in order to obtain the upper bound (which is related to the constraint on the production of dust) and the lower bound (which is related to the cosmological singularity) of the scale factor, we need to presuppose specific forms of $\sigma_m(t)$ and $\Gamma$. We take $\sigma_m(t)=(p\,t)^{\frac23}$ ($p>0$ to guarantee that the entropy of the system is always increasing) and $\Gamma=\frac{g}{t}$ as an example to illustrate the issues in detail. In this case, inequality~(\ref{342}) can be expressed as
	\begin{eqnarray}\label{346}
		p\,t\leq \left(\frac{2\pi k c}{\hbar}\right)^{3/2}\left [\frac32(G_0\rho_0)^{\frac12}a_0^{\frac32}t_0^{-\frac{g}2}\frac2{g+2}\left(t^{\frac{2+g}2}-t_0^{\frac{2+g}2}\right)+a_0^{\frac32}\right].
	\end{eqnarray}
	For convenience, we can set $\frac{2\pi k c}{\hbar}=1$, $\frac32(G_0\rho_0)^{\frac12}=1$, $a_0=1$, and $t_0=1$. Then, the Bekenstein bound is reduced to
	\begin{eqnarray}\label{347}
		0\leq \frac2{g+2}t^{\frac{2+g}2}+\frac g{g+2}-	p\,t.
	\end{eqnarray}
	
\begin{figure}[h]
		\centering
		\renewcommand{\figurename}{Fig.}
		\includegraphics[width=12cm,height=8cm]{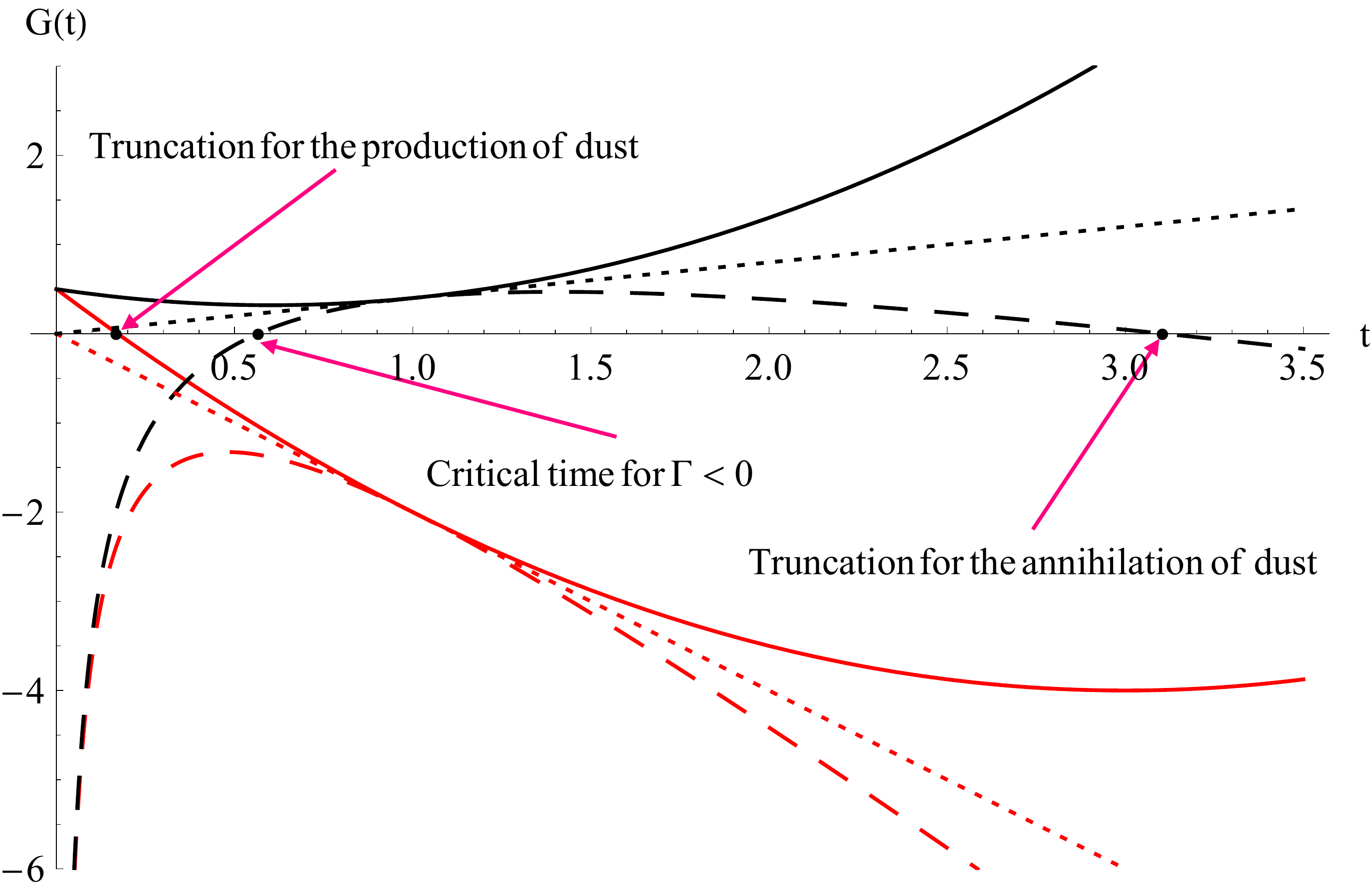}
		\caption{Plot of $G(t)$. There are six sets of the parameters ($g$, $p$): $g=2$ ($\Gamma>0$) and $p=3$ marked with the red solid line; $g=2$ ($\Gamma>0$) and $p=\frac35$ marked with the black solid line; $g=0$ ($\Gamma=0$) and $p=3$ marked with the red dotted line; $g=0$ ($\Gamma=0$) and $p=\frac35$ marked with the black dotted line; $g=-3$ ($\Gamma<0$) and $p=3$ marked with the red dashed line;  $g=-3$ ($\Gamma<0$) and $p=\frac35$ marked with the black dashed line.} \label{F5}		
	\end{figure}

	We can set the right-hand side of the above inequality as a new function $G(t)$, which is plotted in Fig.~\ref{F5} with some representative values of the parameters ($g$, $p$). From Fig.~\ref{F5}, we can judge whether the universe has the initial singularity for different situations and obtain the constraint of the Bekenstein bound on the production of dust. For different values of the parameters ($g$, $p$), the results are totally different.
	
	When $g=0$ ($\Gamma=0$), as long as $p\leq1$, the Bekenstein bound is always valid (see inequality~(\ref{347}) and the black dotted line), so it can not avoid the cosmological singularity. For $p>1$, the Bekenstein bound does not hold at any time (see inequality~(\ref{347}) and the red dotted line), so this situation can not happen inside the co-moving volume of the dust-dominated universe.
	
	When $g=2$ ($\Gamma>0$) and $p=3$ (see the red solid line), since the Bekenstein bound holds at the beginning ($t=0$) of the universe, it can not avoid the cosmological singularity. Note that for $\Gamma>0$ the scale factor is naturally nonzero at the beginning ($t=0$) of the universe. So, there is still no initial singularity and the Bekenstein bound can not modify the critical scale of the universe. But over time, the production of dust will break the Bekenstein bound, so it requires a truncation for the interaction between dust and other matter (or the background space-time). When $g=2$ ($\Gamma>0$) and $p=\frac35$ (see the black solid line), the Bekenstein bound is tenable. Therefore, it can not avoid the cosmological singularity and constrain the production of dust.
	
	When $g=-3$ ($\Gamma<0$) and $p=3$ (see the red dashed line), the Bekenstein bound is invalid at any time, so it is an unreasonable case. When $g=-3$ ($\Gamma<0$) and $p=\frac35$ (see the black dashed line), there exists a nonzero critical time. However, a nonzero critical time does not signify a nonzero critical scale of the universe. For $\Gamma<0$, according to the solution of the scale factor (see Fig.~\ref{F1}), only when the nonzero critical time is larger than the minimum value of time, the cosmological singularity can be avoided. One can take the nonzero critical time into the solution of the scale factor and then it is found that the scale factor is negative, which means that the Bekenstein bound can not prevent the appearance of the cosmological singularity. With the increase of $t$, there will be a new intersection between the black dashed line and the horizontal axis, which requires the annihilation of dust to be cut off in the late universe.
	
	Here we can briefly summarize the above discussion as follows. When $\Gamma\leq0$, the Bekenstein bound can not avoid the cosmological singularity, but there could exist a truncation for the annihilation of dust. When $\Gamma>0$, there is no cosmological singularity due to the solution of the scale factor, and the Bekenstein bound can not affect the critical value of the scale factor (the critical scale of the universe). As for whether there is a truncation for the production of dust, it depends on the values of the parameters ($g$, $p$). All of these results are based on $\sigma_m(t)=(p\,t)^{\frac23}$ ($p>0$) and $\Gamma=\frac{g}{t}$. For other forms of $\sigma_m(t)$ and $\Gamma$, the results could be completely different.
	
\begin{figure}[h]
		\centering
		\renewcommand{\figurename}{Fig.}
		\includegraphics[width=12cm,height=8cm]{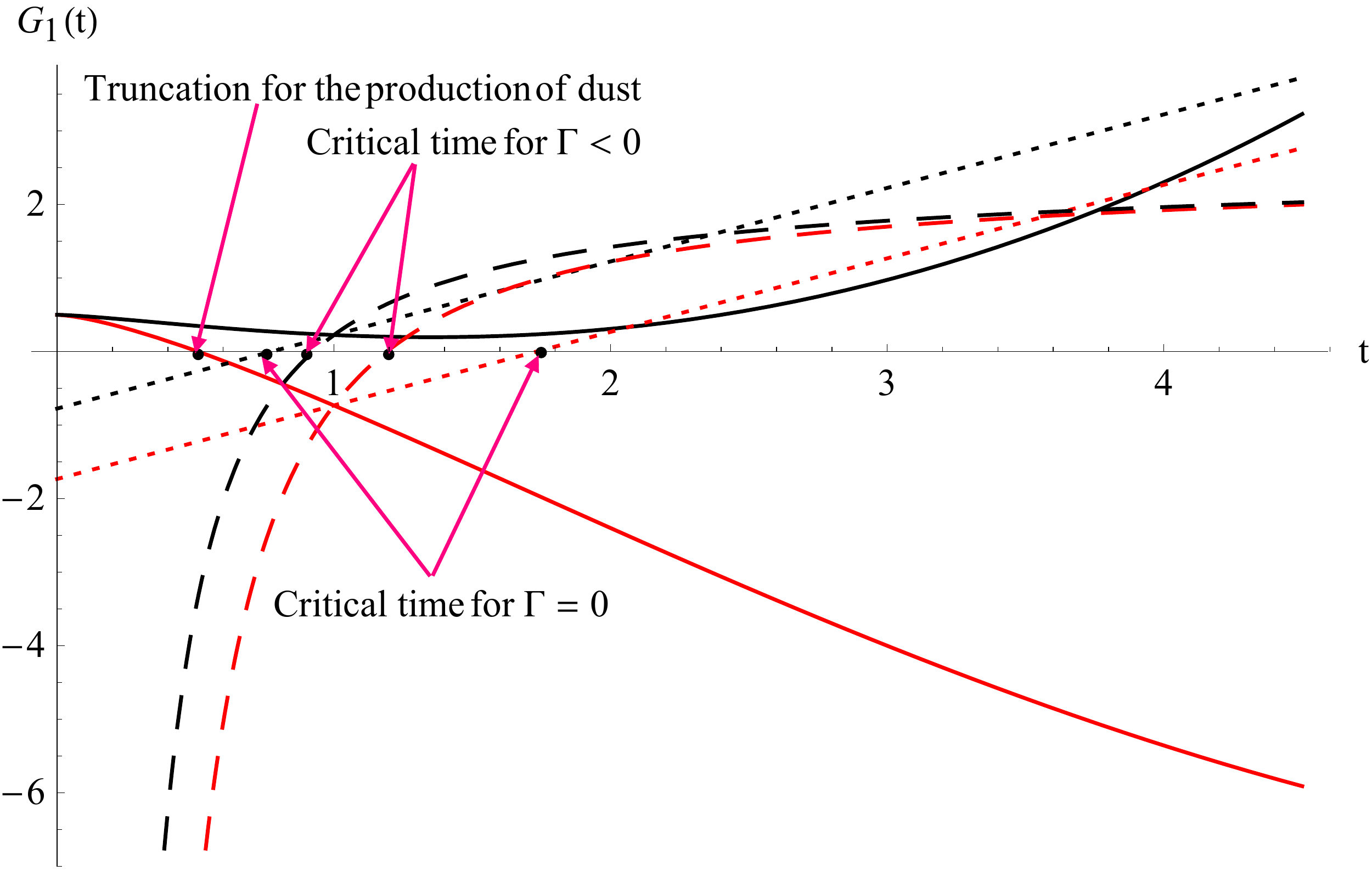}
		\caption{Plot of $G_1(t)$. There are six sets of the parameters ($g$, $p$): $g=2$ ($\Gamma>0$) and $p=3$ marked with the red solid line; $g=2$ ($\Gamma>0$) and $p=\frac35$ marked with the black solid line; $g=0$ ($\Gamma=0$) and $p=3$ marked with the red dotted line; $g=0$ ($\Gamma=0$) and $p=\frac35$ marked with the black dotted line; $g=-3$ ($\Gamma<0$) and $p=3$ marked with the red dashed line;  $g=-3$ ($\Gamma<0$) and $p=\frac35$ marked with the black dashed line.} \label{F6}		
	\end{figure}

	If we use the spherical entropy bound to discuss the cosmological singularity and the constraint on the production of dust inside the co-moving volume, we need the entropy of dust inside the co-moving volume to satisfy
	\begin{eqnarray}\label{348}
		S=\sigma(t)N_0\,\text{exp}\left[\int_{t_0}^{t}\Gamma\,\text d t \right]\leq \frac{k a^2}{4l_p^2}.
	\end{eqnarray}
	We can still set $\sigma(t)=(p\,t)^{\frac23}$ ($p>0$) and $\Gamma=\frac{g}{t}$. Note that the spherical entropy bound does not involve the equivalent mass of a single particle, so we do not have to define the entropy of dust per unit mass. The spherical entropy bound can be expressed as
	\begin{eqnarray}\label{349}
		(p\,t)^{2/3}N_0\left(\frac t {t_0}\right)^g\leq \frac{k}{4l_p^2}\left [\frac32(G_0\rho_0)^{\frac12}a_0^{\frac32}t_0^{-\frac{g}2}\frac2{g+2}\left(t^{\frac{2+g}2}-t_0^{\frac{2+g}2}\right)+a_0^{\frac32}\right]^{4/3}.
	\end{eqnarray}
	For convenience, we can set  $\frac{k}{4l_p^2}=1$, $\frac32(G_0\rho_0)^{\frac12}=1$, $N_0=1$, $a_0=1$, and $t_0=1$. Then, it could be further simplified as
	\begin{eqnarray}\label{350}
		0\leq \frac2{g+2}t^{\frac{2+g}2}+\frac g{g+2}-p^{\frac12} t^{\frac{3g}{4}}.
	\end{eqnarray}
	Similarly, we set the right-hand side of the above inequality as a new function $G_1(t)$, which is plotted in Fig.~\ref{F6} with some representative values of the parameters ($g$, $p$).
	
	Comparing Fig.~\ref{F6} with Fig.~\ref{F5}, one can find that when $\Gamma>0$ (see the solid lines), the results are similar. Therefore, we will not rehash the discussion on the case. When $\Gamma=0$ (see the dotted lines), there exists a nonzero critical time ($0<p^{1/2}\leq t_c$), which can be obtained directly from inequality~(\ref{350}) with $g=0$. Since the evolution of the scale factor starts from $a(t=0)=0$ for $\Gamma=0$, the cosmological singularity can be avoided, which is different from the previous case (see the dotted lines in Fig.~\ref{F5}). Apparently, as the parameter $p$ grows, the critical time will be larger and so the critical scale of the universe will be larger. When $\Gamma<0$ (see the dashed lines), there exists a nonzero critical time. But, the nonzero critical time usually can not avoid the cosmological singularity because $a(t=0)<0$ for $\Gamma<0$. Only when the critical time is larger than the minimum value of time, the cosmological singularity can be avoided. In addition, there does not exist a truncation for the annihilation of dust. Again, these results are based on $\sigma(t)=(p\,t)^{\frac23}$ ($p>0$) and $\Gamma=\frac{g}{t}$. For other forms of $\sigma(t)$ and $\Gamma$, one may get different results.
	
	\subsection{Particle horizon}
	\label{sec42}
\begin{figure}[h]
		\centering
		\renewcommand{\figurename}{Fig.}
		\includegraphics[width=8cm]{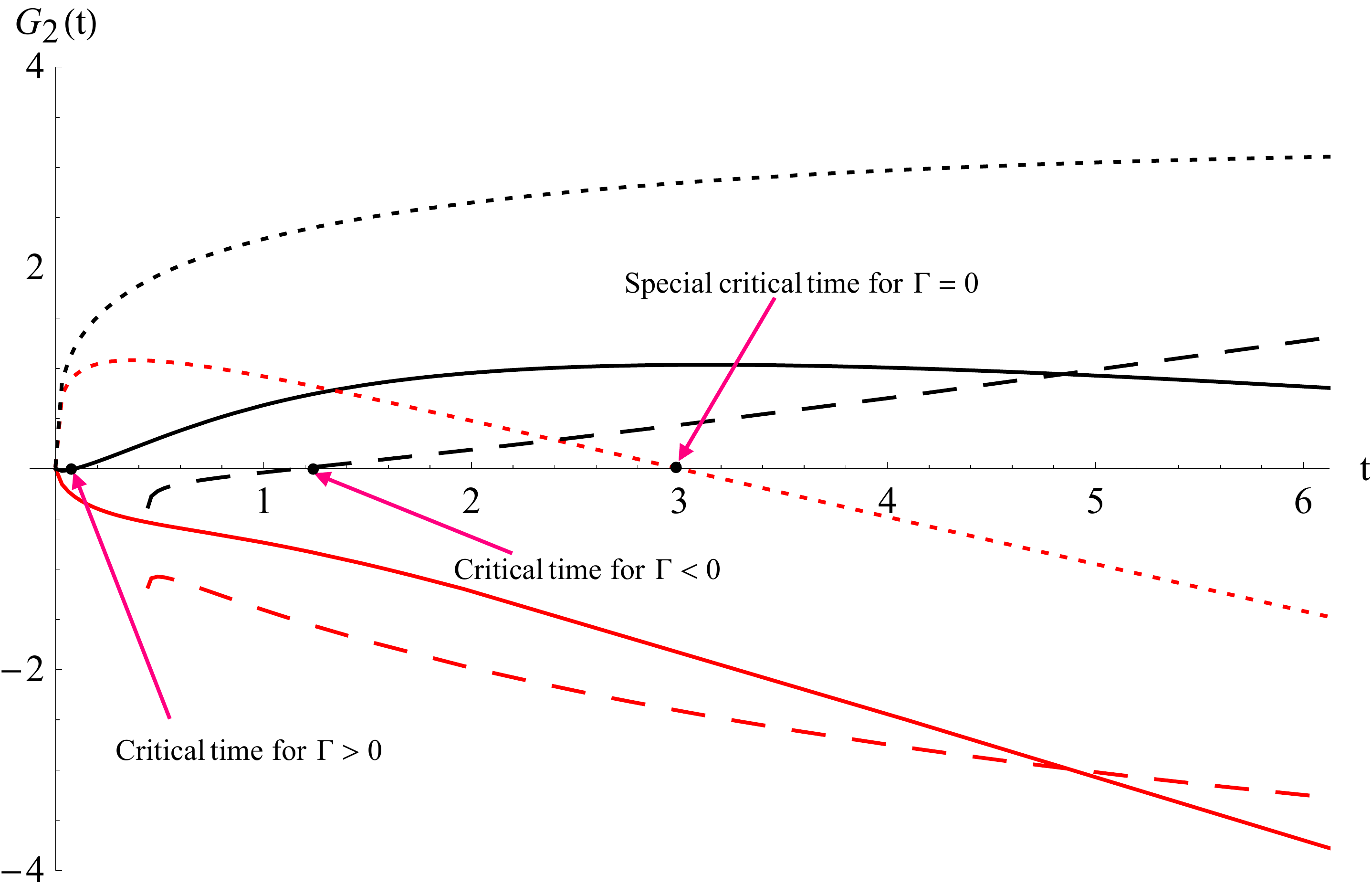}
		\includegraphics[width=8cm]{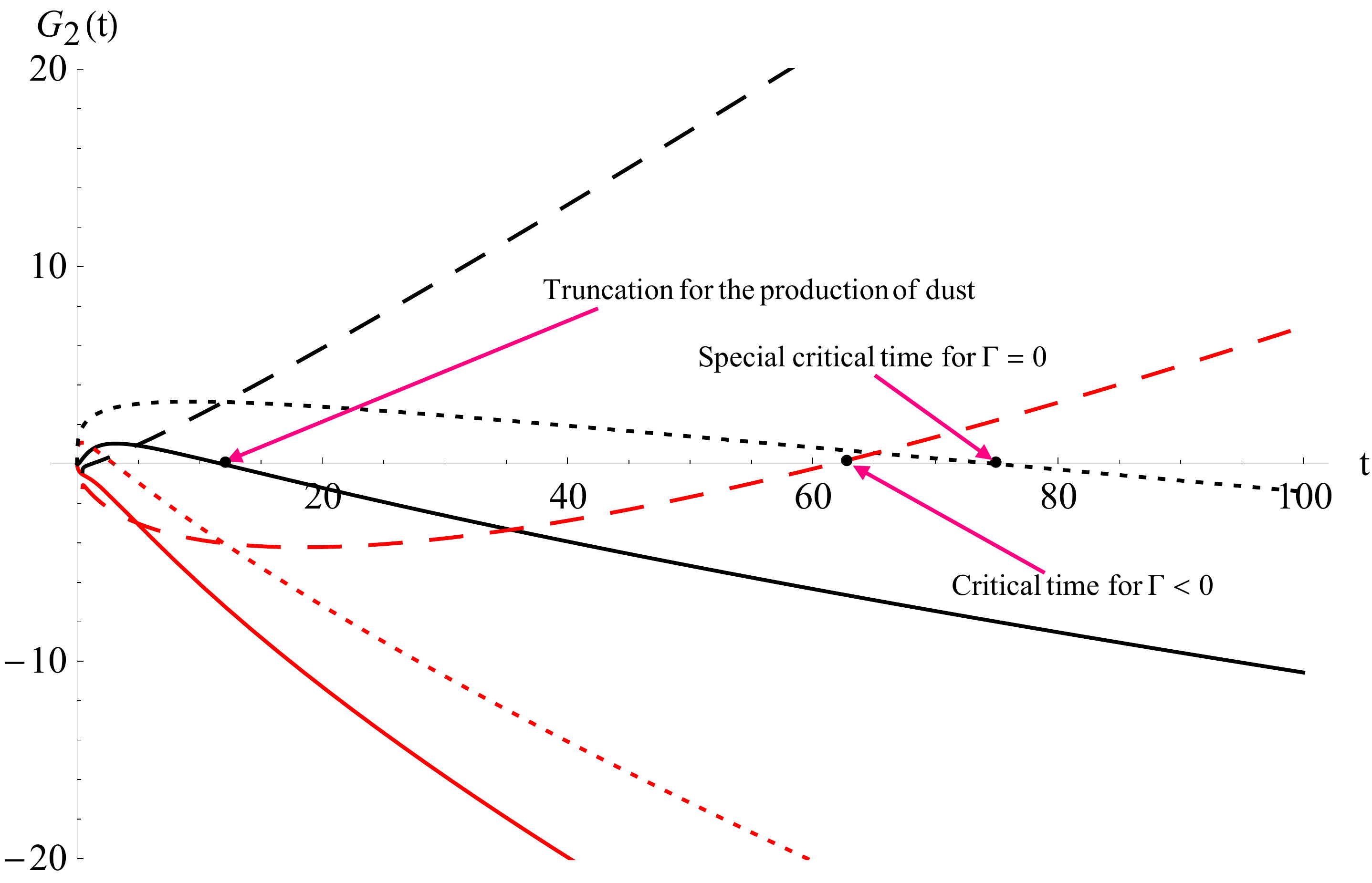}
		\caption{Plot of $G_2(t)$. There are six sets of the parameters ($g$, $p$): $g=2$ ($\Gamma>0$) and $p=3$ marked with the red solid line; $g=2$ ($\Gamma>0$) and $p=\frac35$ marked with the black solid line;	$g=0$ ($\Gamma=0$) and $p=3$ marked with the red dotted line; $g=0$ ($\Gamma=0$) and $p=\frac35$ marked with the black dotted line; $g=0$ ($\Gamma=0$); $g=-3$ ($\Gamma<0$) and $p=3$ marked with the red dashed line;  $g=-3$ ($\Gamma<0$) and $p=\frac35$ marked with the black dashed line.} \label{F7}		
	\end{figure}

	At last, we consider the volume covered by the particle horizon of a given observer in the dust-dominated universe. The solution of the scale factor is also given by Eq.~(\ref{345}). We still set $\Gamma=\frac gt$, $c=1$, $\frac32(G_0\rho_0)^{\frac12}=1$, $a_0=1$, and $t_0=1$. So, the particle horizon can be expressed as
	\begin{eqnarray}\label{351}
		R_H\!\!\!&=&\!\!\!\frac{(g+2)t}{g}  \sqrt[3]{\frac{2 t^{\frac{g}{2}+1}+g}{g+2}} \, _2F_1\left(1,\frac{g+8}{3 g+6};\frac{g+4}{g+2};-\frac{2 t^{\frac{g}{2}+1}}{g}\right)\nonumber\\
		\!\!\!&-&\!\!\!\frac{(g+2)t_s}{g}  \sqrt[3]{\frac{2 t_s^{\frac{g}{2}+1}+g}{g+2}} \, _2F_1\left(1,\frac{g+8}{3 g+6};\frac{g+4}{g+2};-\frac{2 t_s^{\frac{g}{2}+1}}{g}\right),
	\end{eqnarray}
	where $_2F_1(a,b,c,z)$ is a hypergeometric function. According to our previous assumption, for any $\Gamma$, if $F(t)=\int\frac{c}{a(t)}\text d t$ and $F(t_s)=0$, then the Bekenstein bound (replacing the scale factor in inequality~(\ref{342}) with $R_H$) can be expressed as
	\begin{eqnarray}\label{352}
		0\leq  \frac{(g+2)t}{g}  \sqrt[3]{\frac{2 t^{\frac{g}{2}+1}+g}{g+2}} \, _2F_1\left(1,\frac{g+8}{3 g+6};\frac{g+4}{g+2};-\frac{2 t^{\frac{g}{2}+1}}{g}\right)-(p\,t)^{\frac23}.
	\end{eqnarray}
	Here, we have set $\frac{2\pi k c}{\hbar }=1$ and $\sigma_m(t)=(p\,t)^{\frac23}$ ($p>0$). The right-hand side of the above inequality is defined as a new function $G_2(t)$, which is plotted in Fig.~\ref{F7} with some representative values of the parameters ($g$, $p$). In all cases, $t_s$ is the corresponding minimum value of time.
	
	From Fig.~\ref{F7}, we can find that when $\Gamma>0$ (see the solid lines), there are two possible consequences. For a smaller $p$ (such as $g=2$ and $p=\frac35$, the black solid line), there exists a truncation for the production of dust due to the Bekenstein bound (see the black solid line in the right panel), and the cosmological singularity can be avoided by the Bekenstein bound (see the black solid line in the left panel). In this case, the Bekenstein bound will make the critical scale of the universe bigger. For a larger $p$ (such as $g=2$ and $p=3$, the red solid line), the Bekenstein bound can not be satisfied, so this situation can not happen inside the volume covered by the particle horizon. When $\Gamma=0$ (see the dotted lines), there is only one possible consequence. The entropy of dust inside the volume covered by the particle horizon is consistent with the Bekenstein bound at the beginning ($t=0$) of the universe, and then it will break the Bekenstein bound at some point (i.e., the special critical time). Therefore, in this case, the Bekenstein bound can not avoid the cosmological singularity but it provides an upper bound for time (i.e., the special critical time). In this work, we can not offer a definitive explanation for the upper bound of time. When $\Gamma<0$ (see the dashed lines), it is found that the critical time is larger than the minimum value ($t_s=\frac49$) of time (at which the scale factor is vanishing). Therefore, the cosmological singularity can be avoided. Comparing the critical time for $p=3$ (see the red dashed line in the right panel) and the one for $p=\frac35$(see the black dashed line in the left panel), one can find that as $p$ increases, the critical scale of the universe will be larger. In addition, we can find that the annihilation of dust can not be truncated in the late universe.

\begin{figure}[h]
		\centering
		\renewcommand{\figurename}{Fig.}
		\includegraphics[width=8cm]{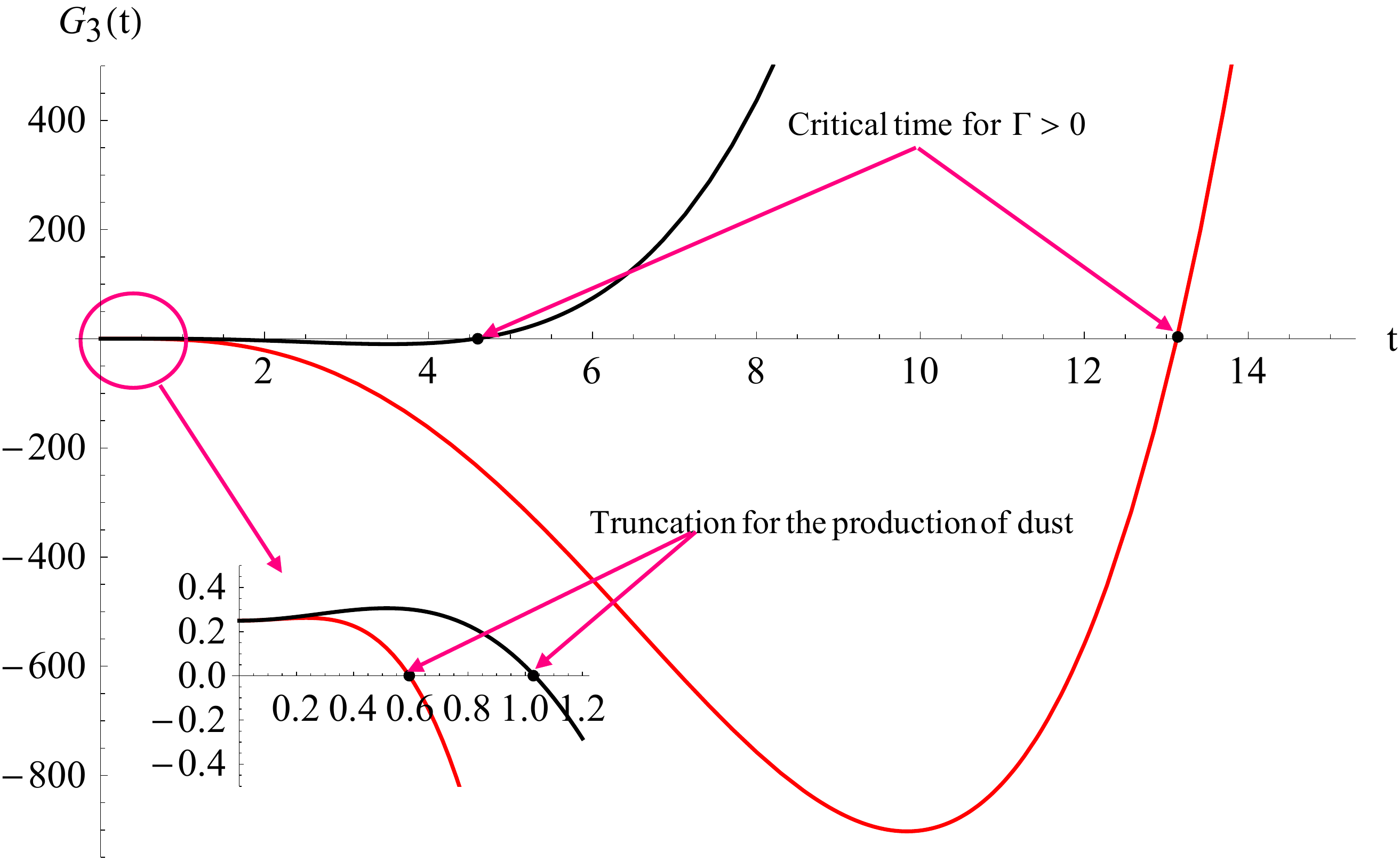}
		\includegraphics[width=8cm]{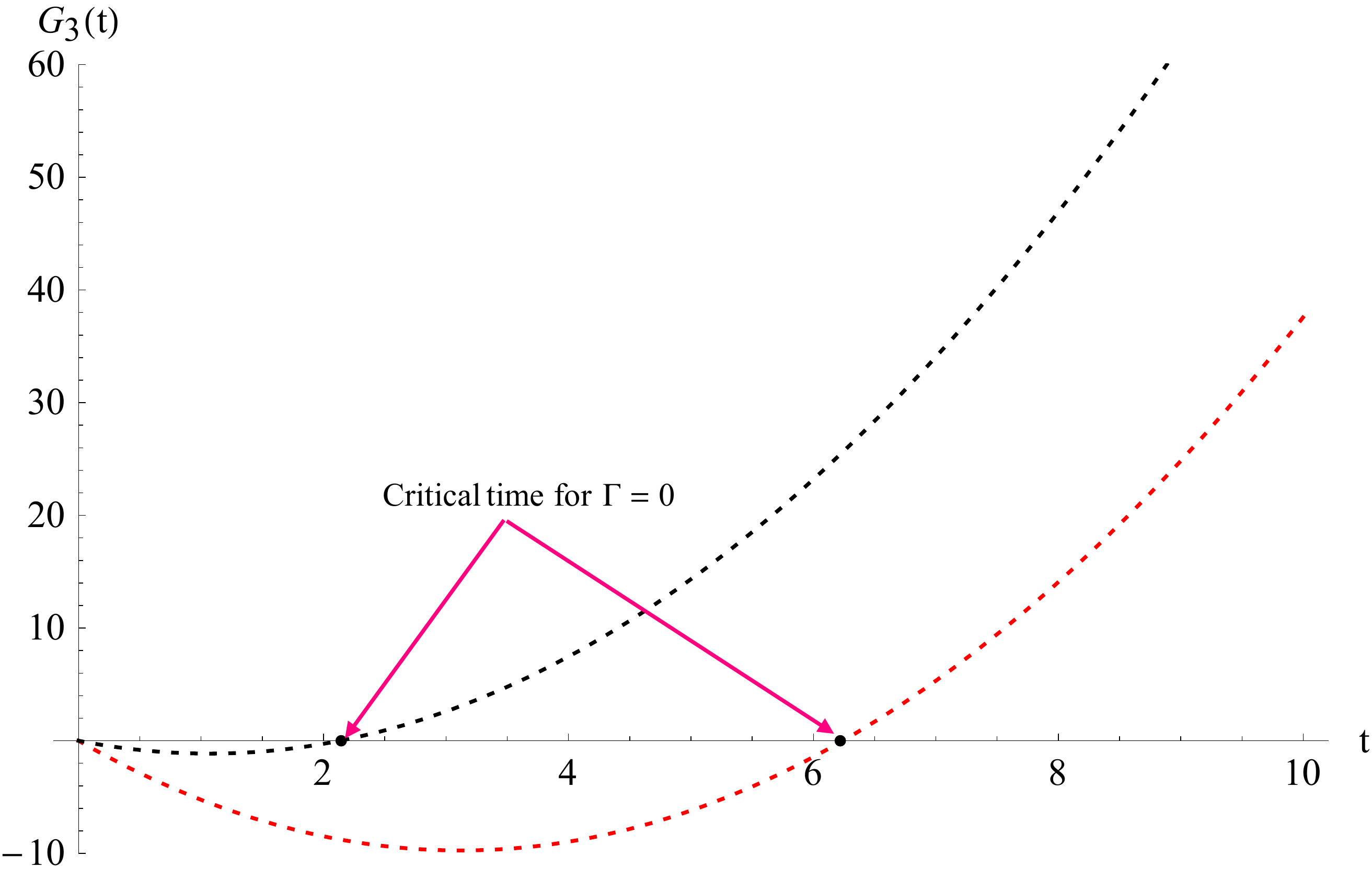}
		\includegraphics[width=8cm]{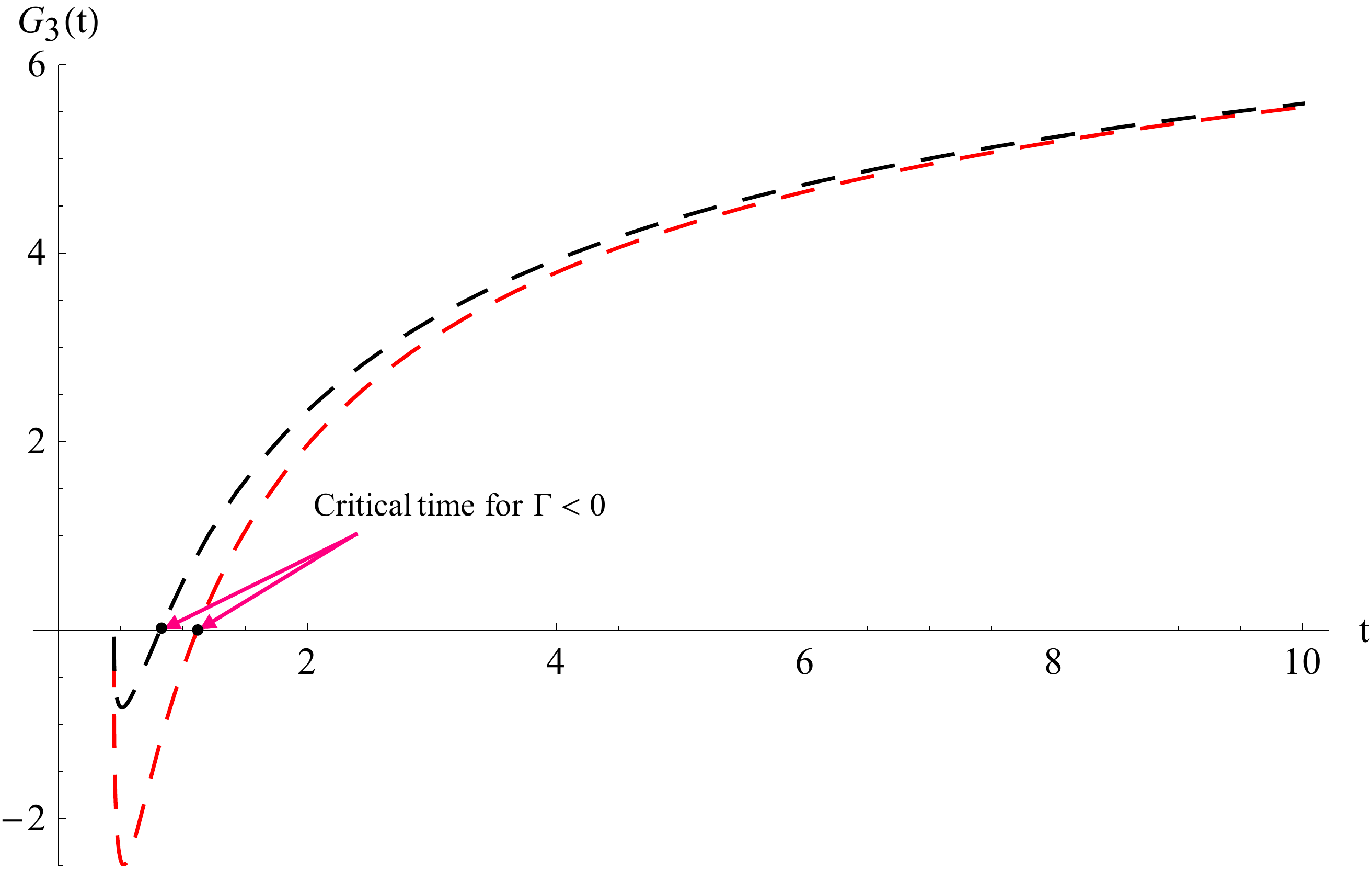}
		\caption{Plot of $G_3(t)$. There are six sets of the parameters ($g$, $p$). The upper left panel corresponds to $g=2$ ($\Gamma>0$), where $p=3$ is marked with the red solid line and $p=\frac35$ is marked with the black solid line. The upper right panel corresponds to $g=0$ ($\Gamma=0$), where $p=3$ is marked with the red dotted line and $p=\frac35$ is marked with the black dotted line. The lower panel corresponds to $g=-3$ ($\Gamma<0$), where $p=3$ is marked with the red dashed line and $p=\frac35$ is marked with the black dashed line.} \label{F8}		
	\end{figure}

For the spherical entropy bound, the entropy of dust inside the volume covered by the particle horizon of a given observer needs to satisfy
	\begin{eqnarray}\label{353}
		S=\sigma(t)N_0\,\text{exp}\left[\int_{t_0}^{t}\Gamma\,\text d t \right] \frac{R_H^3}{a^3}\leq \frac{k R_H^2}{4l_p^2}.
	\end{eqnarray}
	In this case, we do not need to define the entropy of dust per unit mass. Note that since $\Gamma$ is the particle production rate of dust inside the co-moving volume, there is an extra dimensionless factor $\frac{R_H^3}{a^3}$ on the left-hand side of the above inequality. 	Similarly, we set $\sigma(t)=(p\,t)^{\frac23}$ ($p>0$) and $\Gamma=\frac{g}{t}$. With the previous parameter settings, the spherical entropy bound can be reduced as
	\begin{eqnarray}\label{354}
		0\leq \left(\frac2{g+2}t^{\frac{2+g}2}+\frac g{g+2}\right)^2-(p\,t)^{\frac23}\cdot t^{g}\cdot\frac{(g+2)t}{g}  \sqrt[3]{\frac{2 t^{\frac{g}{2}+1}+g}{g+2}} \, _2F_1\left(1,\frac{g+8}{3 g+6};\frac{g+4}{g+2};-\frac{2 t^{\frac{g}{2}+1}}{g}\right) .
	\end{eqnarray}
	The right-hand side of the above inequality is defined as a new function $G_3(t)$, which is plotted in Fig.~\ref{F8} with some representative values of the parameters ($g$, $p$). In all cases, $t_s$ is still the corresponding minimum value of time.
	
	 From the upper left panel in Fig.~\ref{F8}, it can be found that when $\Gamma>0$, there exists a critical time and a truncation for the production of dust due to the spherical entropy bound. But the critical time is later than the time corresponding to the truncation, which seems strange. Here, we can deem that the truncation is non-physical and the universe began at the critical time. In this case, the universe has no initial singularity and the production of dust can not be cut off in the late universe. But, if we accept the existence of the truncation, then the universe originated from the minimum value ($t=0$) of time, but has no initial singularity due to the solution of the scale factor. Therefore, the spherical entropy bound can not modify the critical scale of the universe. The production of dust can only occur in the early universe and beyond the critical time. According to the upper right panel and lower panel in Fig.~\ref{F8}, when $\Gamma\leq0$, the cosmological singularity can be avoided by the spherical entropy bound. And there is no truncation for the annihilation of dust.
	
	  Finally, we emphasize again that these results are based on the specific assumptions about the model and the parameters. The results may be completely different for other models and parameters.

	\section{Conclusions and Discussions}
	\label{sec5}
	
	The Bekenstein bound has been proposed more than 30 years, after which there emerge multiple definitions of entropy bounds. Although most of these entropy bounds are established on black hole research, their applications in cosmology are also widely studied, especially the topics related to the initial singularity and entropy of the universe. The entropy of the universe caused by particle production is usually constrained by the (generalized) second law of thermodynamics and the law of thermal equilibrium, but the results in practice are not satisfactory~\cite{Yu:2018qzl,SolaPeracaula:2019kfm}. In this work, we use entropy bounds (the Bekenstein bound and the spherical entropy bound) to constrain the entropy of particles in two toy cosmological models with particle production, thereby constraining the production of the corresponding particles. We also study the effect of particle production on the cosmological singularity in light of entropy bounds.
	
	For the two toy cosmological models we study, the cosmological singularity can be avoided by considering entropy bounds in some special cases, such as the system covered by the particle horizon in the radiation-dominated universe with $\Gamma>0$ constrained by the Bekenstein bound (see Fig.~\ref{F2}). In addition, the production (annihilation) of particles needs to be truncated in some special cases due to entropy bounds, such as the co-moving volume in the dust-dominated universe with $\Gamma<0$ and $p=\frac35$ constrained by the Bekenstein bound (see Fig.~\ref{F5}). There are also some cases that always do not satisfy certain entropy bound, and thus they can not happen in the corresponding cosmological model, such as the system covered by the particle horizon in the dust-dominated universe with $\Gamma>0$ and $p=3$ constrained by the Bekenstein bound (see Fig.~\ref{F7}). Moreover, some cases always satisfy certain entropy bound, which means that the corresponding entropy bound does not play much of a role, such as the co-moving volume in the dust-dominated universe with $\Gamma>0$ and $p=\frac35$ constrained by the spherical entropy bound (see Fig.~\ref{F6}). In general, for different situations there are various possible results about the cosmological singularity and the truncation of particle production (annihilation). It should be emphasized that these results are mainly dependent on the selections of the system in the universe and the entropy bound.
	
	Since most entropy bounds are obtained from black hole research, they may not apply to the universe. Therefore, it is worth discussing whether we can employ directly the entropy bound obtained from black hole research to cosmology. Seeking an entropy bound which is really practical in cosmology is the focus of our next research, which can also avoid the emergence of various possible results for different entropy bounds. To sum up, our methods are basically appropriate for all cosmological models with particle production, but it is still a preliminary attempt. How to implement the researches in the real universe and how to choose the system in the universe and the entropy bound need to be further studied in the future.

	\section*{Acknowledgments} \hspace{5mm}
	This work was supported by the National Natural Science Foundation of China (Grants No. 11873001, and No. 12047564) and the Postdoctoral Science Foundation of Chongqing (Grant No. cstc2021jcyj-bsh0124).

	%%%%%%%%%%%%%%%%%%%%%%%%%%%%%%%%
\end{document}